\documentclass[twocolumn,english,aps,prb,floats,showpacs]{revtex4}
\usepackage[T1]{fontenc}
\usepackage[latin9]{inputenc}
\usepackage{babel}
\usepackage{bm}
\usepackage{amsmath}
\usepackage{amssymb}
\usepackage{graphicx}
\usepackage{wasysym}
\usepackage{esint}
\usepackage[unicode=true,
 bookmarks=false,
 breaklinks=false,pdfborder={0 0 1},backref=section,colorlinks=false]
 {hyperref}
\usepackage{breakurl}

\makeatletter
\@ifundefined{textcolor}{}
{%
 \definecolor{BLACK}{gray}{0}
 \definecolor{WHITE}{gray}{1}
 \definecolor{RED}{rgb}{1,0,0}
 \definecolor{GREEN}{rgb}{0,1,0}
 \definecolor{BLUE}{rgb}{0,0,1}
 \definecolor{CYAN}{cmyk}{1,0,0,0}
 \definecolor{MAGENTA}{cmyk}{0,1,0,0}
 \definecolor{YELLOW}{cmyk}{0,0,1,0}
}

\@ifundefined{textcolor}{}{%
 \definecolor{BLACK}{gray}{0}
 \definecolor{WHITE}{gray}{1}
 \definecolor{RED}{rgb}{1,0,0}
 \definecolor{GREEN}{rgb}{0,1,0}
 \definecolor{BLUE}{rgb}{0,0,1}
 \definecolor{CYAN}{cmyk}{1,0,0,0}
 \definecolor{MAGENTA}{cmyk}{0,1,0,0}
 \definecolor{YELLOW}{cmyk}{0,0,1,0}
}

\usepackage{babel}
\usepackage{ifpdf}\usepackage{bm}

\makeatother

\begin{document}

\title{Skyrmion-Skyrmion Interaction in a Magnetic Film}

\author{Daniel Capic, Dmitry A. Garanin, and Eugene M. Chudnovsky}

\affiliation{Physics Department, Herbert H. Lehman College and Graduate School,
The City University of New York, 250 Bedford Park Boulevard West,
Bronx, New York 10468-1589, USA }

\date{\today}
\begin{abstract}
Interaction of two skyrmions stabilized by the ferromagnetic exchange, Dzyaloshinskii-Moriya interaction (DMI), and external magnetic field has been studied numerically on a 2D lattice of size large compared to the separation, $d$, between the skyrmions. We show that two skyrmions of the same chirality (determined by the symmetry of the crystal) repel. In accordance with earlier analytical results, their long-range pair interaction falls out with the separation as $\exp(-d/\delta_H)$, where $\delta_H$ is the magnetic screening length, independent of the DMI. The prefactor in this expression depends on the DMI that drives the repulsion. The latter results in the spiral motion of the two skyrmions around each other, with the separation between them growing logarithmically with time. When two skyrmions of the total topological charge $Q = 2$ are pushed close to each other, the discreteness of the atomic lattice makes them collapse into one skyrmion of charge $Q = 1$ below a critical separation. Experiment is proposed that would allow one to measure the interaction between two skyrmions by holding them in positions with two magnetic tips. Our findings should be of value for designing topologically protected magnetic memory based upon skyrmions.
\end{abstract}

\pacs{}
\maketitle

\section{Introduction}

Magnetic skyrmions are topological defects of the uniform ferromagnetic background in two dimensions (2D). They have been initially introduced in field models of elementary particles and atomic nuclei \cite{SkyrmePRC58,Polyakov-book,Manton-book,D1,D2}, and later entered condensed matter physics in applications to 2D ferromagnets and antiferromagnets \cite{BelPolJETP75,Lectures}, Bose-Einstein condensates \cite{AlkStoNat01}, quantum Hall effect \cite{SonKarKivPRB93,StonePRB93}, anomalous Hall effect \cite{YeKimPRL99}, and liquid crystals \cite{WriMerRMR89}. Unlike micron-size magnetic bubbles studied in the past \cite{MS-bubbles,ODell}, skyrmions in magnetic films can be small compared to the domain wall thickness, making them good candidates for extra dense magnetic memory. Currently they represent a very active field of research due to their potential for topologically-protected nanoscale information processing \cite{Nagaosa2013,Zhang2015,Klaui2016,Leonov-NJP2016,Hoffmann-PhysRep2017,Fert-Nature2017}.

Theoretical research on magnetic skyrmions has been focused on contribution of various interactions to skyrmion structure and dynamics. Skyrmions can be stabilized by the magnetic field in systems with Dzyaloshinskii-Moriya interaction (DMI) that are lacking inversion symmetry \cite{Bogdanov94,Bogdanov-Nature2006,Heinze-Nature2011,Boulle-NatNano2016,Leonov-NJP2016} or in materials with quenched randomness \cite{EC-DG-PRL2018,EC-DG-NJP2018} whose effect is similar to that of the DMI. In experiments, individual skyrmions can be created, annihilated and moved by current-induced spin-orbit torques \cite{Yu-NanoLet2016,Fert-Nature2017,Legrand-Nanolet2017}, by pushing elongated magnetic domains through a constriction using an in-plane current \cite{Jiang-Sci2015,Hoffmann-PhysRep2017}, by spin-polarized currents from a scanning tunneling microscope \cite{Romming-Sci2013}, by laser-generated heat pulses \cite{Berruto-PRL2018}, and by cutting stripe magnetic domains with a tip of a scanning magnetic force microscope \cite{Senfu-APL2018}.  They can also be nucleated by temperature \cite{Yu2010,Zhang2018,GC-MMM2019} and written in a film with an MFM tip \cite{AP-2018}.

When skyrmions are formed and moved in experiments they incidentally appear close to each other. The natural question is how they interact. For skyrmion lattices this question is usually swept under the rug by considering overall stability of the skyrmion crystal that cannot be easily traced to the pair interaction between  skyrmions. The separate problem of skyrmion-skyrmion pair interaction has been addressed by a few authors but the results have been unclear and at times contradictory. Lin et al. \cite{Lin2013} used Thiele's approach \cite{Thiele} to derive particle-like equations of motion for skyrmions. They found that skyrmions repel with the force that goes down with their separation $d$ as $K_1(d/\xi)$ where $K_1$ is the Macdonald function and $\xi$ is a characteristic length describing the overlap of the spin fields of the two skyrmions.  Lin and Hayami \cite{Lin2016} further investigated this problem within effective Ginzburg-Landau theory. In their model the interaction depended on the relative helicity of the two skyrmions and behaved non-monotonically on the separation, oscillating between repulsion and attraction. R\'{o}zsa et al.  \cite{Rozsa2016} demonstrated that frustration of the exchange interaction can lead to the attractive interaction between skyrmions at short distances. Attractive magnetic skyrmions have been predicted by Leonov et al. \cite{Leonov2016} and observed by Loudon et al. \cite{Loudon2018} in the cone phase of a cubic helimagnet Cu$_2$OSeO$_3$. Evidence of the switching of skyrmion-skyrmion interaction from attraction to repulsion on increasing magnetic field, obtained with the help of high-resolution Lorentz transmission electron microscopy, has been reported by Du et al.  \cite{Du2018} in B20-type FeGe nanostripes.

The subtleness of the problem can be seen from the following argument. In the generic BP model, a skyrmion of arbitrary dimension with arbitrary topological charge $Q$ is stable. It applies in particular to a biskyrmion with $Q = 2$ that consists of two skyrmions of size $\lambda$ separated by distance $d$. The energy of the BP biskyrmion is independent of $\lambda$ and $d$, which corresponds to zero interaction between the two skyrmions in the biskyrmion. This somewhat counterintuitive mathematical fact implies that interactions other than exchange must be responsible for skyrmion-skyrmion interaction. Stable biskyrmion lattices have been reported in centrosymmetric magnetic films \cite{Yu-2014,Zhang2016,Loudon2019,Yao2019,Gobel2019}. It was shown theoretically \cite{BL2019} that such lattices can be stabilized by the perpendicular magnetic anisotropy and dipole-dipole interaction. Individual biskyrmions can be stable in the presence of Bloch lines \cite{GCZ-EPL2017} or frustrated Heisenberg exchange \cite{XZhang2017,Leonov-NatCom2015}. Metastable biskyrmion configurations have been observed in the Landau-Lifshitz dynamics of a frustrated bilayer film \cite{Nowak2017}. Stability of a biskyrmion at a certain separation, $d$, of two skyrmions of opposite chirality, that occurs in a nonchiral film in the narrow field range, has been demonstrated in Ref.\ \onlinecite{Capic2019}. For such a biskyrmion it implies repulsion between skyrmions at small distances and attraction at large distances.

In this paper we study skyrmion-skyrmion interaction in the practically relevant ferromagnetic model with the DMI and external magnetic field. It provides stable $Q = 1$ skyrmions of unique chirality determined by the symmetry of the DMI.  The size of the skyrmion is controlled by the magnetic field. We compute numerically the energy of two such skyrmions as function of their separation on a spin lattice of dimensions that are much greater than the skyrmion size. That energy minus twice the energy of an individual non-interacting skyrmion is interpreted as the energy of the skyrmion-skyrmion interaction. We find it to be a repulsion that with excellent accuracy follows the law $\exp(-d/\delta_H)$ at large separations. Here $\delta_H = a\sqrt{J/H}$ is the magnetic screening length, with $a$ being the lattice spacing, $J$ being the exchange constant, and $H$ being the external magnetic field in the energy units.  When one takes into account the scaling of variables used in Ref.\ \onlinecite{Lin2013}, where $d$ was measured in units of $J/A$ (with $A$ being the strength of the DMI), the two results coincide. The non-scaled dependence reveals an interesting fact that the characteristic length of the exponential decay of the interaction on separation is independent of the DMI. However, the numerical results indicate that the prefactor here depends on the DMI that drives the repulsion.

We show that the repulsion between the skyrmions induces their gyroscopic rotational dynamics that is typical for magnetic systems. Skyrmions spiral around each other with their separation increasing logarithmically with time. Rigorous analytical arguments are provided that explain most of our numerical results. We then propose a method of measuring interaction between skyrmions by controlling them with magnetic force microscope (MFM) tips and measuring the interaction between the tips.

The paper is organized as follows. Theory of a biskyrmion in a continuous spin-field model is presented in Section \ref{continuous}. Numerical model and computational results for the skyrmion-skyrmion interaction are described in Section \ref{Fixed_Spins}. Dynamics of interacting skyrmions is discussed in Section \ref{dynamics}. Analytical arguments that explain numerical results are given in Section \ref{theory}. In Section \ref{experiment} we propose experiment on measuring interaction between skyrmions. Our results and their practical implications are summarized in Section \ref{Sec_Discussion}.

\section{Skyrmions and skyrmion pairs in a continuous spin-field model}\label{continuous}

Within the pure exchange model, ground state solutions minimize the exchange energy:
\begin{equation}
E_{ex}= \frac{J}{2} \int dx dy \left[\left(\frac{\partial \mathbf{s}}{\partial x}\right)^2+ \left(\frac{\partial \mathbf{s}}{\partial y}\right)^2\right], \label{Eex}
\end{equation}
Here ${\bf s}(x,y)$ is a unit-length three-component dimensionless spin field in a 2D plane and $J$ is the exchange constant that absorbs the length of the spin. These minimum energy solutions are characterized by the topological charge:
\begin{equation}
Q=\frac{1}{4\pi} \int dx dy  \left(\frac{\partial {\bf s}}{\partial x} \times  \frac{\partial {\bf s}}{\partial y}\right) \cdot {\bf s},\label{Q}
\end{equation}
an integer number that indicates how many times the spin vector circumscribes the full solid angle 4$\pi$  as the position vector covers the whole 2D plane. Energy minimization leads to the condition:
\begin{equation}
\mathbf{s} \times \nabla^2 \mathbf{s} = 0, \label{Lagrange}
\end{equation}
which corresponds to the spins being collinear with the effective exchange field. In 2D, using Eq. (\ref{Lagrange}), this can be expressed as a first order differential equation \cite{Lectures,BelPolJETP75}, which in vector form is \cite{BL2019}:
\begin{equation}
\frac{\partial \mathbf{s}}{\partial x}=\pm \mathbf{s} \times \frac{\partial \mathbf{s}}{\partial y}, \qquad   \frac{\partial \mathbf{s}}{\partial y}=\mp \mathbf{s} \times \frac{\partial \mathbf{s}}{\partial x}. \label{BPRelation}
\end{equation}

It is convenient to switch to the complex function $\omega=\omega(z)$ with $z=x+iy$, that is defined by the equations \cite{BelPolJETP75}
\begin{equation}
s_{x}+ i s_{y}=\frac{2\omega}{|\omega|^2+1}, \quad s_{z}= \frac{|\omega|^2-1}{|\omega|^2+1} \label{Omega}
\end{equation}
satisfying $\mathbf{s}^2=1$.

From Eq.(\ref{BPRelation}), follows \cite{BelPolJETP75,Lectures, BL2019}:
\begin{equation}
\frac{\partial \omega}{\partial x} = \pm i \frac{\partial \omega}{\partial y}, \label{Omega}
\end{equation}
which is the Cauchy-Riemann (CR) condition that is satisfied for any analytic function. Any analytical $\omega$ that satisfies Eq.(\ref{Omega}) is a solution that minimizes the exchange energy from Eq.(\ref{Eex}). One can consider two of the simplest solutions:  $\omega=({\lambda}/{z})^{|Q|}$ and $\omega=({\lambda}/{z^{*}})^{Q}$, where $z^{*} = x-iy$ which correspond to the plus and minus signs in the CR equation respectively. Ref.\ \onlinecite{Capic2019} considers other solutions. The former is the antiskyrmion with topological charge $Q<0$ and the latter is the skyrmion with topological charge $Q>0$. At infinity, one has $s_{z}= -1$, whereas at the center of the skyrmion, $s_{z}= 1$.

\begin{figure}[ht]
\centering{}\includegraphics[width=8cm]{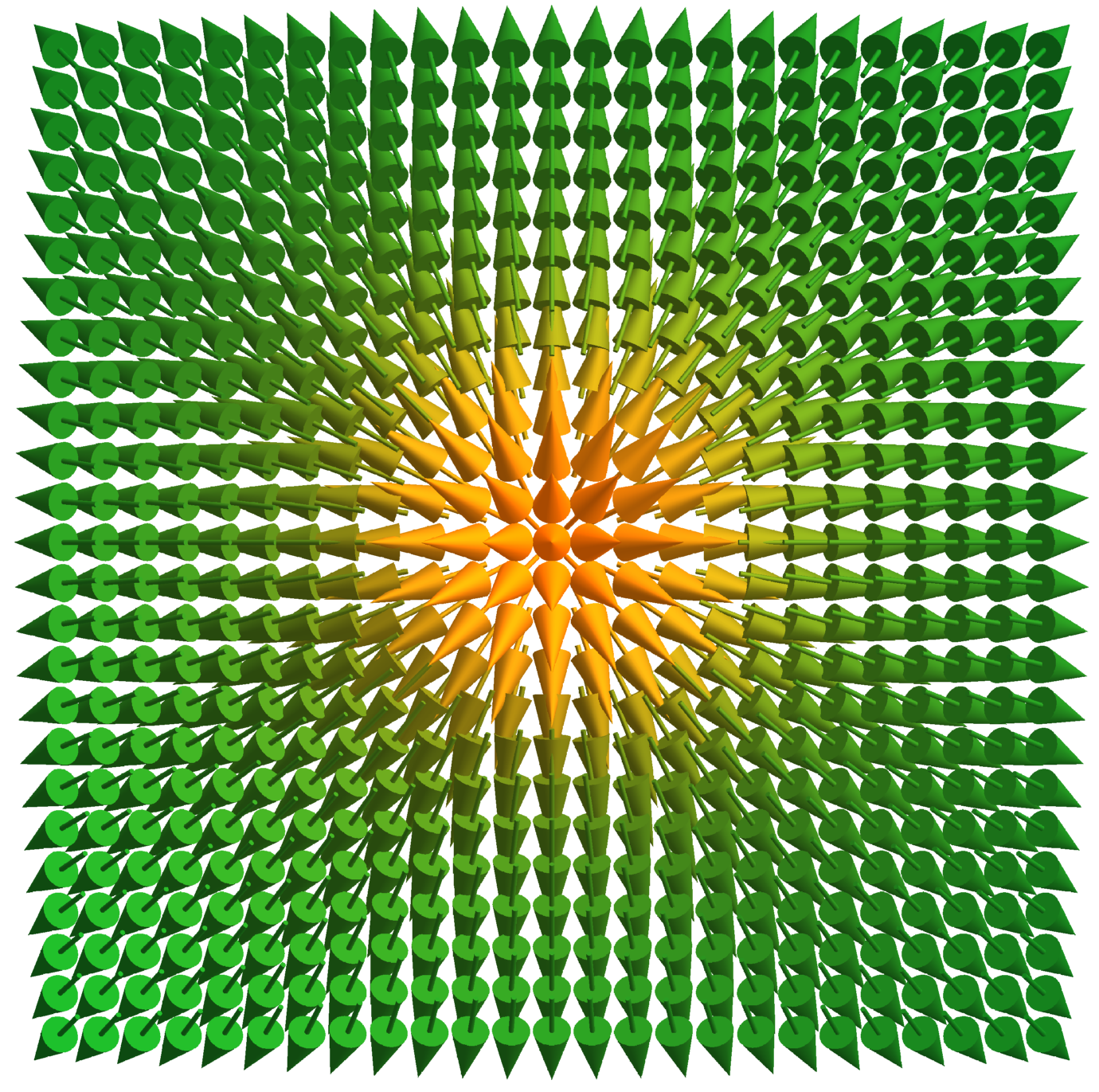} \includegraphics[width=8cm]{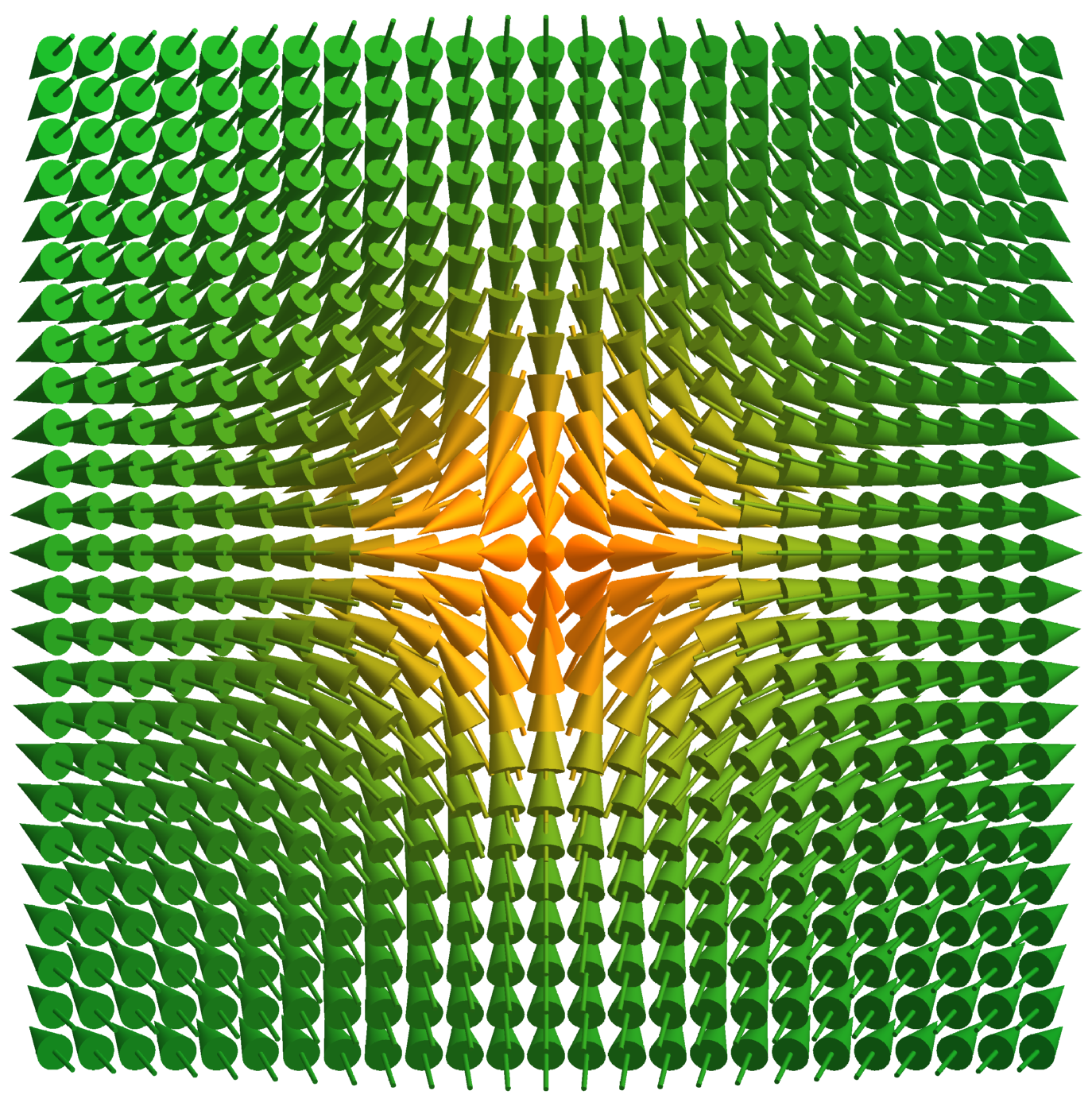}
\caption{N\'{e}el-type skyrmion (upper panel) and antiskyrmion (lower panel) spin-field configurations.}
\label{S-AS}
\end{figure}

The spin components for the $Q=\pm 1$ solution have the form:
\begin{eqnarray}
s_x & = & \frac{2\lambda(x\cos\gamma-y\sin\gamma)}{x^2+y^2+\lambda^2}, \quad s_y = Q\frac{2\lambda(x\sin\gamma+y\cos\gamma)}{x^2+y^2+\lambda^2}, \nonumber \\
s_z & = & -\frac{x^2+y^2-\lambda^2}{x^2+y^2+\lambda^2}. \label{BP}
\end{eqnarray}
The distinction between magnetic skyrmion and antiskyrmion is determined by the different topology of the spin field illustrated in Fig. \ref{S-AS} for $Q = \pm 1$. The N\'{e}el skyrmion and antiskyrmion shown in Fig.\ \ref{S-AS} correspond to $\gamma = 0$.

 A skyrmion of topological charge $Q>0$, defined by Eq.(\ref{Q}) centered at $z=z_{0}$ can be constructed via the choice
\begin{equation}
\omega= \left(\frac{\lambda}{z^{*}-z_{0}^{*}}\right)^{Q} e^{i\gamma},
\end{equation}
where $\lambda$ and $\gamma$ refer to the size and chirality, respectively. A sum of separated poles with different $z_{0}^{*}$ can be used to produce skyrmion lattices \cite{BL2019}.  Specifically for pairs, the $Q=2$ solution with a separation $d$ between what will visually appear as two $Q=1$ skyrmions can be expressed as
\begin{equation}
\omega= \frac{\lambda e^{i\gamma_{1}}}{z^{*}-d/2}+ \frac{\lambda e^{i\gamma_{2}}}{z^{*}+d/2}. \label{OmegaBi}
\end{equation}

The case $\gamma_{1}= \gamma$, $\gamma_{2}= -\gamma$ studied in Ref.\ \onlinecite{Capic2019} is relevant to multilayer centrosymmetric materials. This solution is the opposite chirality BP biskyrmion, where the pair of $Q=1$ skyrmions will have a phase difference of $\pi$ and there is no contribution to the DMI energy. Substitution into Eq.(\ref{OmegaBi}) gives:
\begin{equation}
\omega= \frac{\lambda d e^{i \gamma}}{z^{*2}-(d/2)^2}, \label{OmegaBiopp}
\end{equation}
which in the limit $d \rightarrow 0 $ becomes $\omega= \tilde{\lambda}^2 e^{i\gamma}/z^{*2}$, where $\tilde{\lambda}=\sqrt{\lambda d}$ represents the effective size for small separations. Therefore, there is a continuous transition to $d=0$ that conserves the topological charge $Q=2$.

\begin{figure}[ht]
\centering{}\includegraphics[width=8cm]{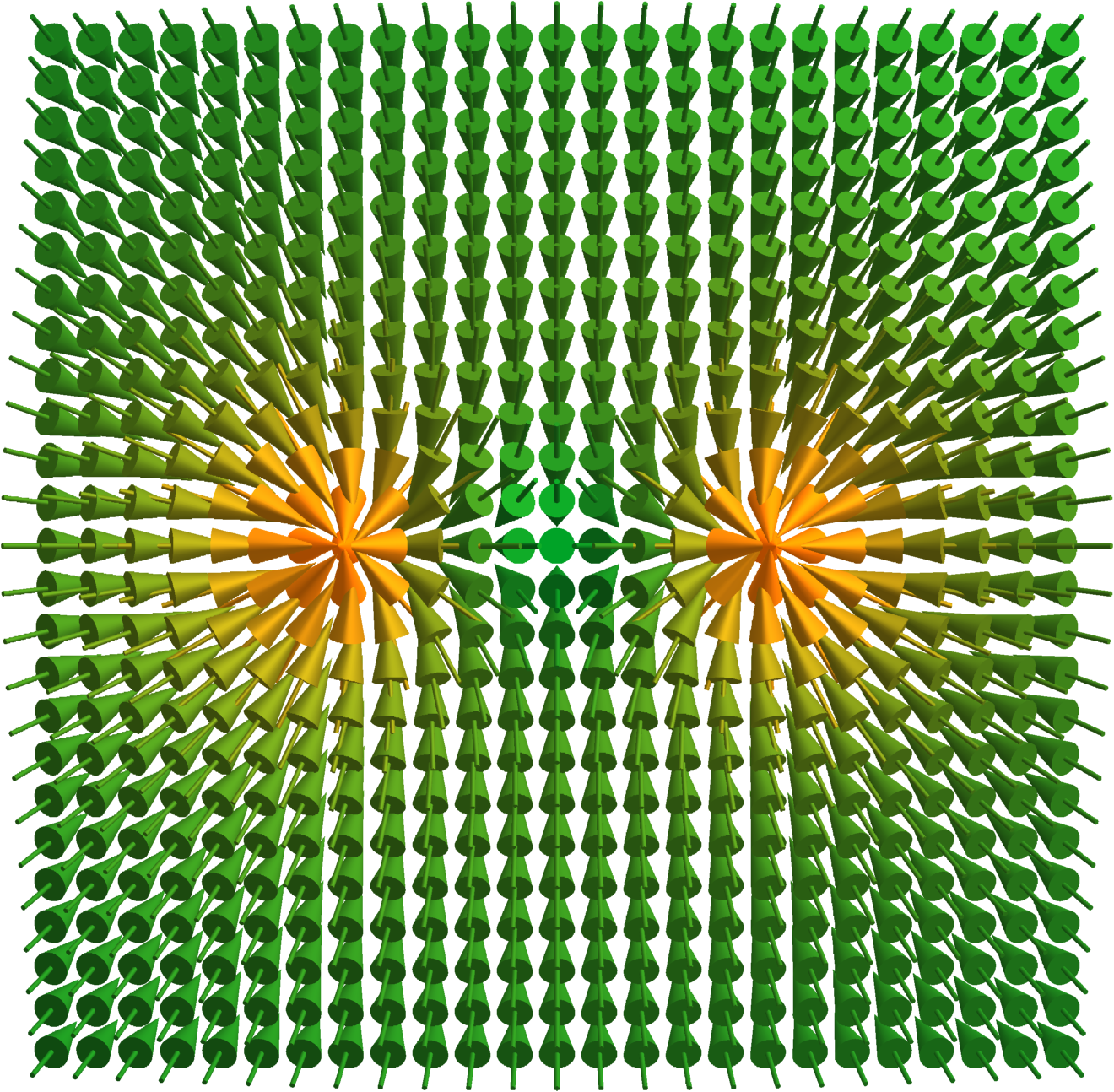} \caption{Spin field in the N\'{e}el-type same chirality BP skyrmion pair.}
\label{BiS}
\end{figure}

In a non-centrosymmetric material the chirality of the skyrmion is determined by the DMI. To study the skyrmion-skyrmion interaction in such a system we must take $\gamma_{1}= \gamma_{2}=\gamma$, which is the same chirality BP skyrmion pair. (We are not using the term biskyrmion here because the two skyrmions are repelling each other and not forming a bound state.) At the center of the same chirality skyrmion pair $s_{z} = -1$, same as at infinity, while at the center of each of the two skyrmions forming the biskyrmion  $s_{z} =1$. From Eq. (\ref{OmegaBi})
\begin{equation}
\omega= \frac{2 \lambda z^{*} e^{i\gamma}}{z^{*2}-(d/2)^2}. \label{OmegaBisame}
\end{equation}
For any finite $d$, the spins point down in the center and $Q=2$. For $d=0$, the topological charge is $Q=1$, which can be seen by setting $d =0$ to obtain $\omega= \tilde{\lambda} e^{i\gamma}/z^{*}$, where $\tilde{\lambda}=2\lambda$. Thus, there is no continuous transition to $d=0$ and instead the system will undergo a topological jump when $d \sim a$, the lattice spacing.  The N\'{e}el  same chirality skyrmion pair with $\gamma = \pi$ is shown in Fig.\ \ref{BiS}.

The spin components of the $Q = 2$ same chirality skyrmion pair are given by:
\begin{small}
\begin{eqnarray}
s_x & = &\frac{16\lambda[(4(x^2+y^2)-d^2)x\cos\gamma-(4(x^2+y^2)+d^2)y\sin\gamma]}{d^4-8d^2(x^2-y^2)+16(x^2+y^2)(x^2+y^2+4\lambda^2)}, \nonumber \\
s_y & = & \frac{16\lambda[(4(x^2+y^2)-d^2)x\sin\gamma+(4(x^2+y^2)+d^2)y\cos\gamma]}{d^4-8d^2(x^2-y^2)+16(x^2+y^2)(x^2+y^2+4\lambda^2)}, \nonumber \\
s_z & = & -\frac{d^4-8d^2(x^2-y^2)+16(x^2+y^2)(x^2+y^2-4\lambda^2)}{d^4-8d^2(x^2-y^2)+16(x^2+y^2)(x^2+y^2+4\lambda^2)}  \label{SpinComponents}
\end{eqnarray}
\end{small}and are used as the initial state for our investigation of the skyrmion-skyrmion interaction.

Using Eq.(\ref{SpinComponents}) as an \textit{Ansatz}, one can compute the energy from the Zeeman and DMI contributions, as we did in Ref. \ \onlinecite{Capic2019} to determine the dependence on the separation $d$. The DMI energy is independent of $d$ and is just double the energy of a single skyrmion. The Zeeman energy only depends logarithmically on $d$. It turns out that the skyrmion-skyrmion interaction is subtle and caused by deviations away from the BP shape that are difficult to compute analytically, which is why we will solve the problem numerically. The results will show a skyrmion-skyrmion repulsion driven by the DMI.

\section{Skyrmion-Skyrmion Interaction in the Fixed Center-Spin Model} \label{Fixed_Spins}

In the numerical work, we consider a lattice model for a 2D ferromagnetic film  with the energy:
\begin{eqnarray}
{\cal H} & = &- \frac{1}{2} \sum_{i j}J_{ij}\mathbf{s}_{i}\cdot \mathbf{s}_{j}-H\sum_{i} s_{iz} \nonumber \\
& - & A\sum_{i} \left[(\mathbf{s}_{i}\times \mathbf{s}_{i+\delta_{x}})_{y} - (\mathbf{s}_{i}\times \mathbf{s}_{i+\delta_{y}})_{x}\right]. \label{Hamiltonian}
\end{eqnarray}
The exchange coupling $J$ is for the nearest neighbors on a square lattice. The stabilizing field $H$ is applied in the negative (downward) \textit{z}-direction to prevent skyrmion expansion, with a magnitude smaller than the collapse field. The DMI term favors outward N\'{e}el-type skyrmions ($\gamma=0$) for $A>0$. Furthermore,  $\delta_{x}$ and $\delta_{y}$ refer to the next nearest lattice site in the positive $x$ or $y$ direction.

The numerical method [\  \onlinecite{Garanin2013} ] involves rotations of the individual spins $\mathbf{s}_{i}$ towards the direction of the local effective field $\mathbf{H}_{\mathrm{eff},i}= -\partial{\cal H} / \partial{\mathbf{s}_{i}}$ with the probability $\alpha$ and the energy-conserving spin flips (\textit{overrelaxation}), $\mathbf{s}_{i} \rightarrow 2(\mathbf{s}_{i} \cdot \mathbf{H}_{\mathrm{eff}, i})\mathbf{H}_{\mathrm{eff}, i}/H_{\mathrm{eff}, i}^2 -\mathbf{s}_{i}$ with the probability $1-\alpha$. The parameter $\alpha$ plays the role of the effective relaxation constant. We use the value $\alpha=0.03$ for the overall fastest convergence.

The system was studied on a square lattice of $800 \times 800$ spins. For most of the work, we use $A/J = -0.01$, with minus signs indicating  materials favoring inward N\'{e}el-type skyrmions, as well as $H/J=-0.00045$ for which $\delta_H \approx 47a$. For the lattice spacing $a$ and the Exchange constant $J$ we chose $a = J = 1$ in the computations. The same-chirality biskyrmion described by the initial BP spin configuration of Eq.\ (\ref{SpinComponents}) was allowed to relax to the minimum-energy  configuration.

Our initial observation was that in the course of the energy minimization, the skyrmions will move away from each other, indicating a repulsive interaction. Separation at which they stop as the repulsion becomes weak can be due to factors, such as, e.g., pinning by the discrete lattice sites. To study the interaction more systematically, we used the fixed center-spin (FCS) method.

To fix the separation between the skyrmions, a strong fictitious field was applied in the positive \textit{z}-direction at two lattice sites separated by the distance $d$ to fix the skyrmion centers  $s_{z}=1$ at these points. The corresponding exchange, Zeeman and DMI energies were recorded for different separations. The separation $d$ was decreased by moving the fixed spins of each skyrmion one lattice spacing closer per one step. The skyrmion size $\lambda$ was determined numerically using the formula \cite{Cai2012}:
\begin{equation}
\lambda_{n}^2= \frac{n-1}{2^{n+1} \pi}a^2\sum_{i}( s_{i,z}+1)^n, \label{Lambda}
\end{equation}
For the computations, we took $\lambda=\lambda_{4}$. Here $\lambda$ provides an estimate of the size of one skyrmion in the pair. The extra factor of 1/2 appearing in Eq.(\ref{Lambda}) is due to the topological charge $Q=2$.

\begin{figure}[ht]
\centering
\begin{minipage}{.5\textwidth}
	\centering
	\includegraphics[width=9cm]{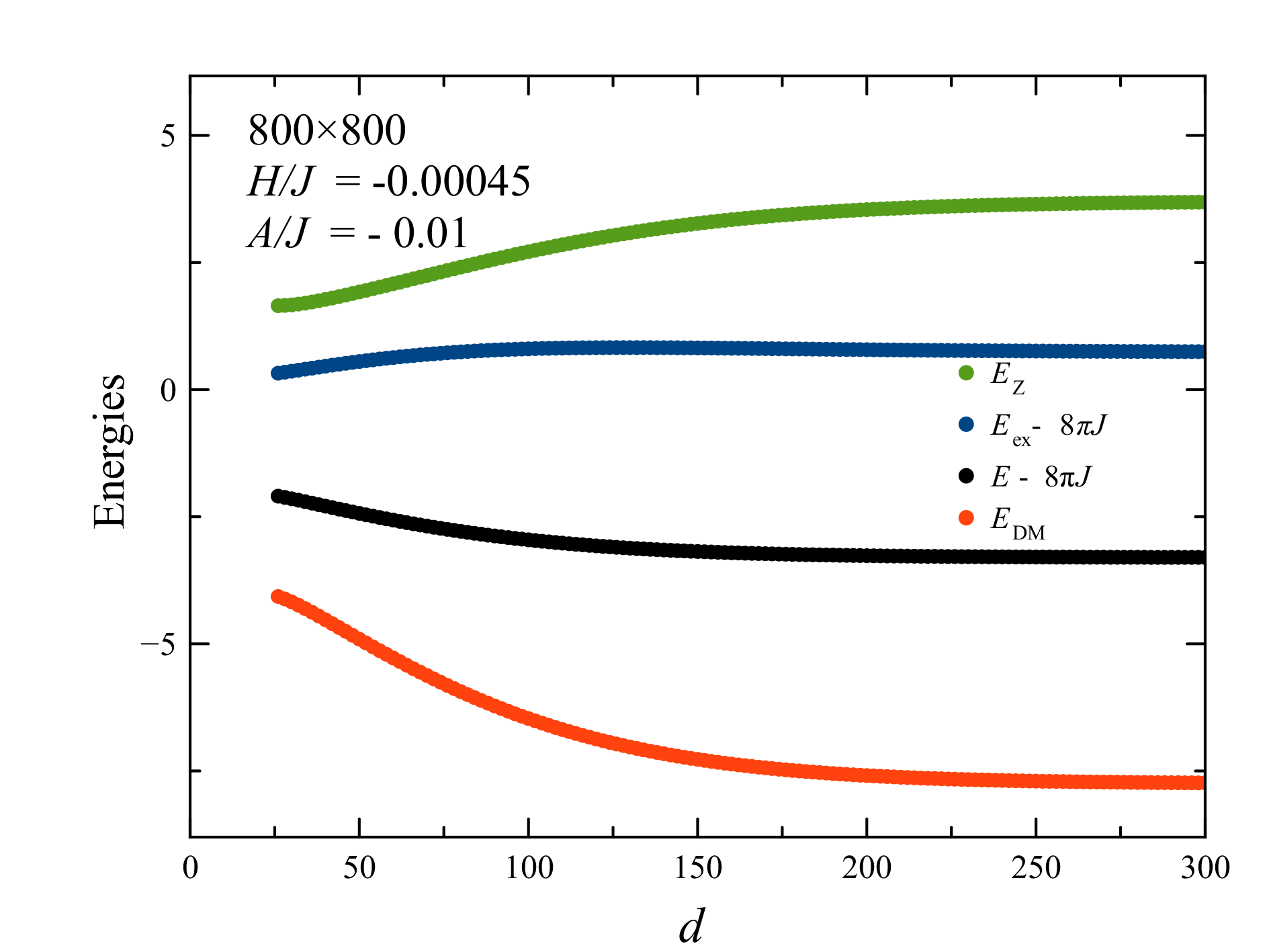}
	\caption{Behavior of the exchange, Zeeman, DMI  and total energies as a function of the separation $d$.}
	\label{AllEnergy}
\end{minipage}
\begin{minipage}{.5\textwidth}
	\centering
	\includegraphics[width=9cm]{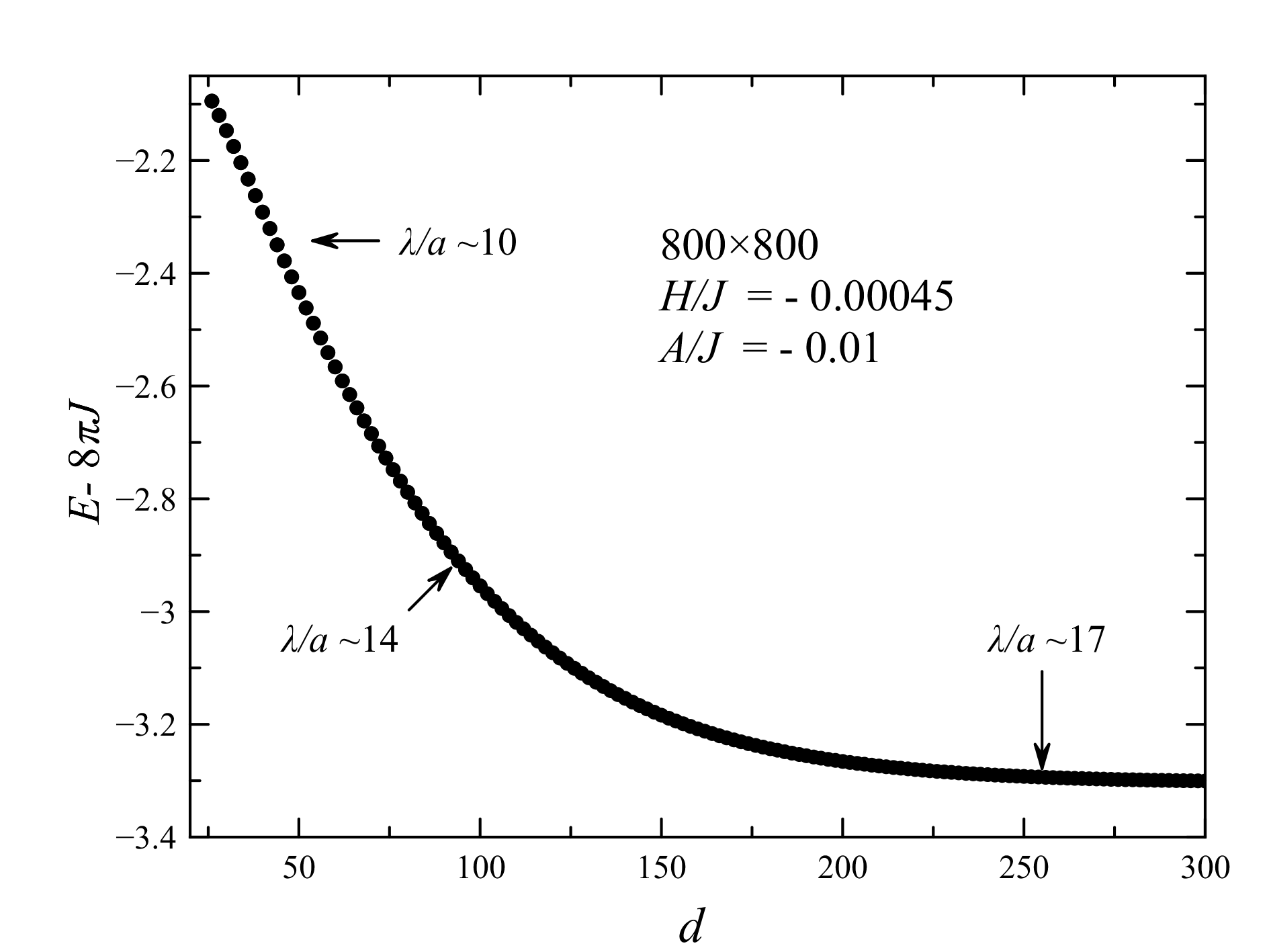}
	\caption{Total energy as a function of the separation $d$.}
\end{minipage}
\end{figure}

The total energy of the system was governed by the interplay between the Zeeman and DMI interactions, while the exchange energy was approximately constant. The skyrmions shrank as they got closer together, leading to a decrease in the (positive) Zeeman energy and an increase in the (negative) DM energy. The net effect was an increase in the total energy, which was lower than $8\pi J$, as the separation decreased, indicative of the repulsion for all $d$. The Zeeman, DMI, exchange and total energies are plotted in Fig. \ref{AllEnergy}. One can see that the skyrmion repulsion is driven by the DMI that is the only interaction whose energy decreases with $d$. This is in contrast with the analytical model that assumed a rigid size $\lambda$.

\begin{figure}[h]
\hspace{-0.5cm}
\centering
\includegraphics[width=9cm]{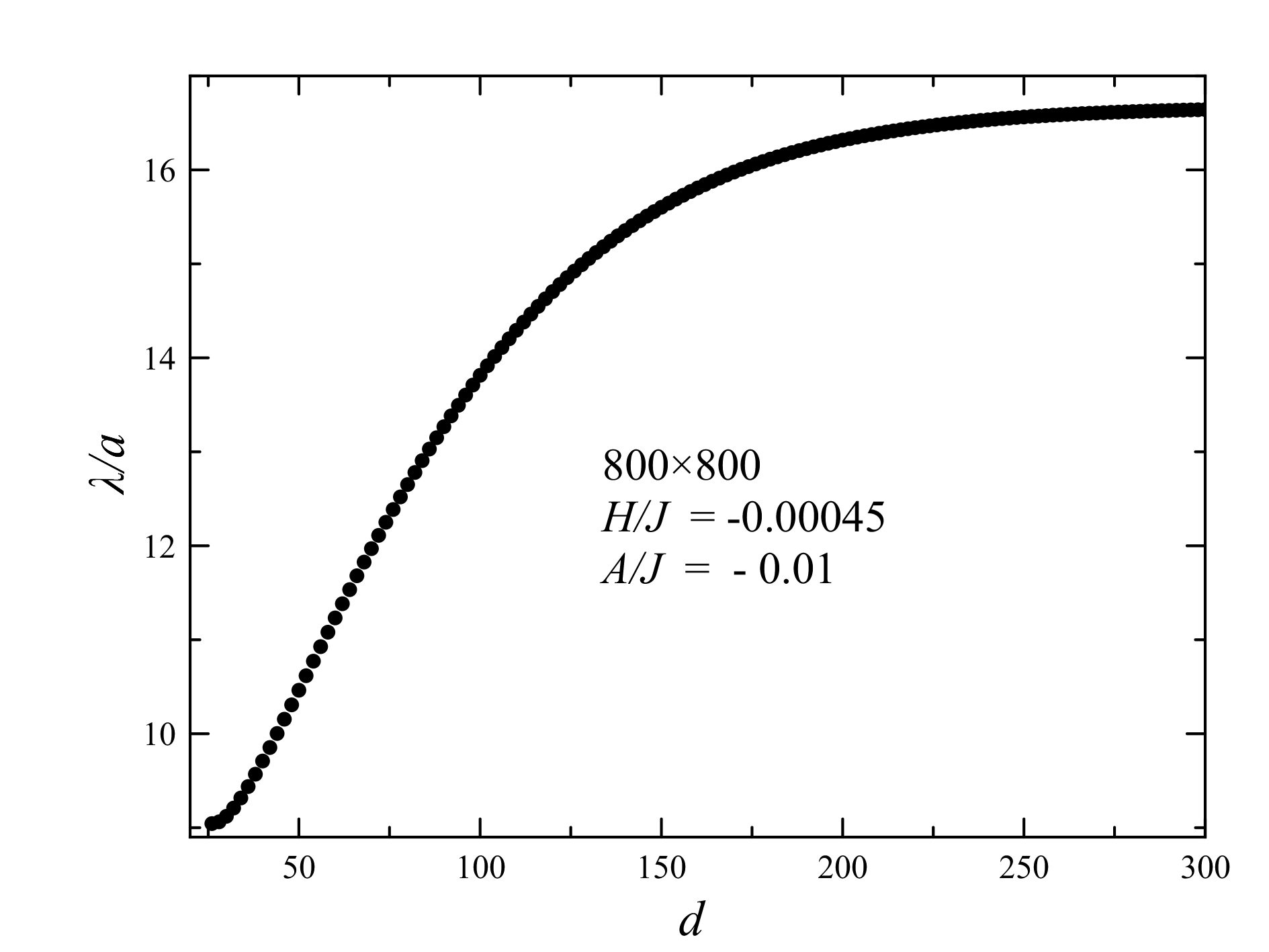}
\caption{Numerical data on the size $\lambda$ as a function of the separation $d$.}
\label{lam}
\end{figure}

 Numerical results have clearly demonstrated correlation between $\lambda$ and $d$, that is, correlation between the size and separation of the interacting skyrmions, see Fig.\ref{lam} .

\begin{figure}[ht]
\hspace{-0.5cm}
\centering
\includegraphics[width=9cm]{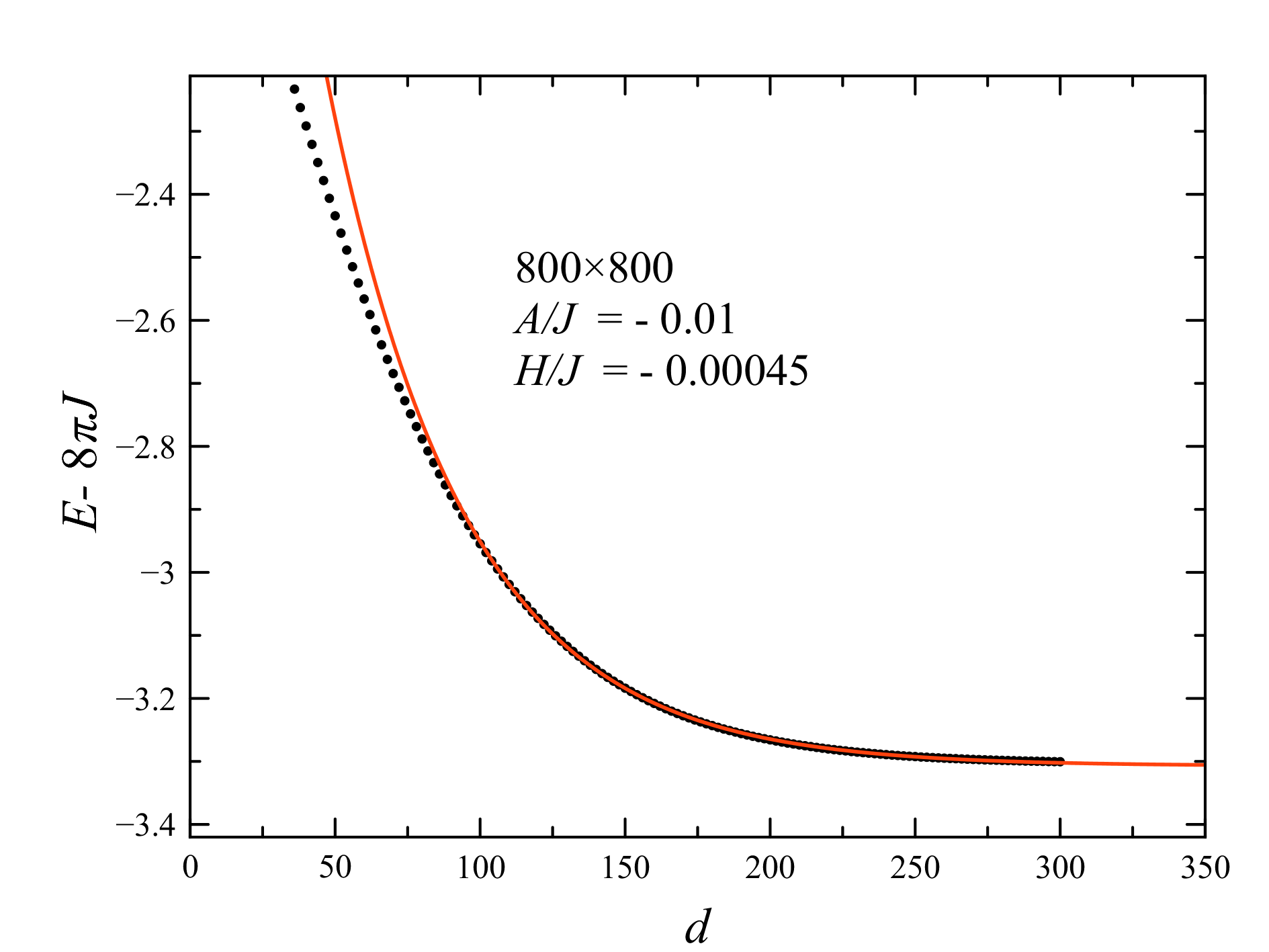}
\caption{Numerical data on the long-range behavior of the total energy fitted by Eq.\ (\ref{LRBehavior}).}
\label{LREnergy}
\end{figure}

\begin{figure}[ht]
\centering
\hspace{-0.5cm}
\includegraphics[width=9cm]{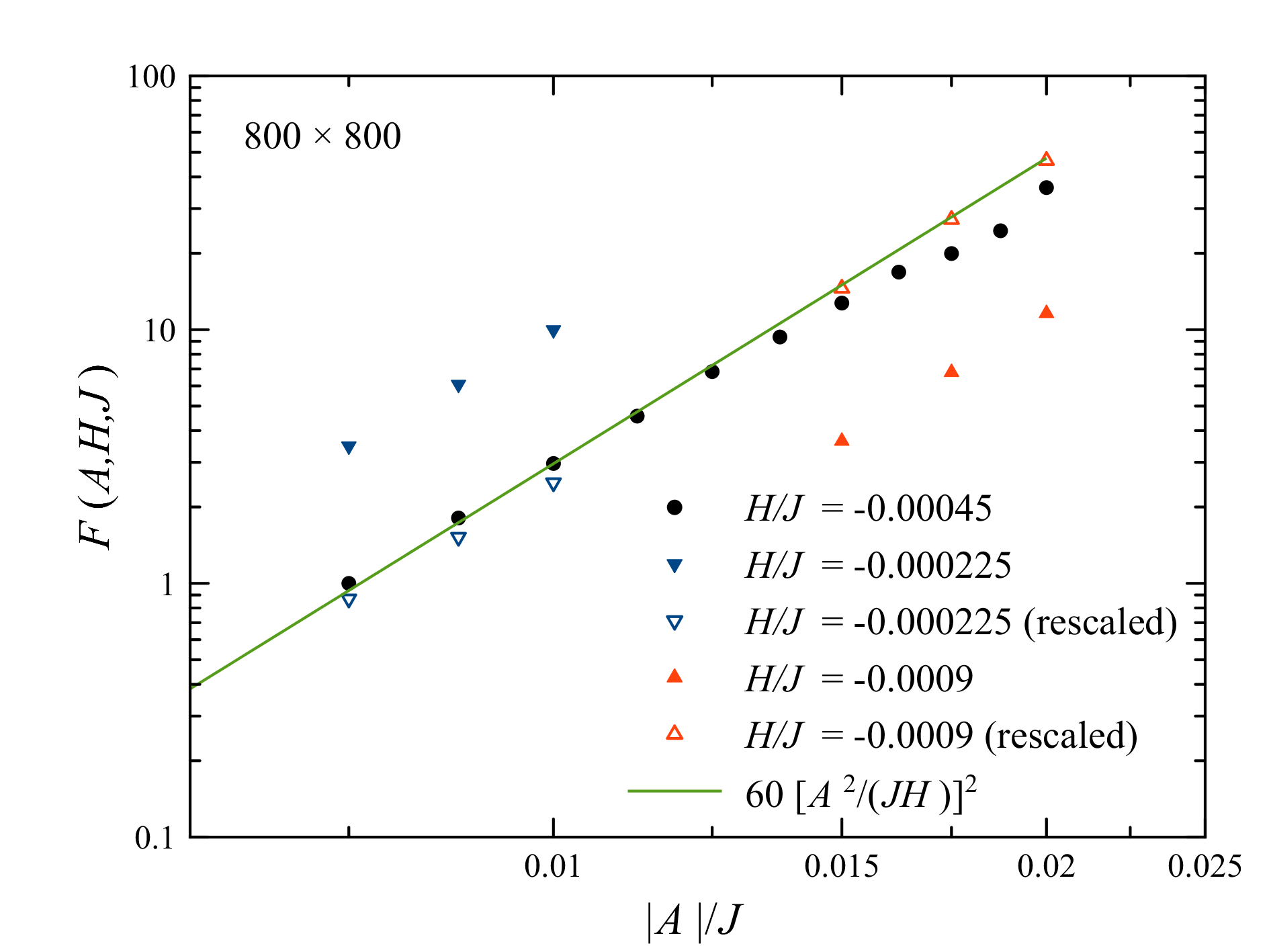}
\caption{Dependence of the prefactor in Eq.\ (\ref{LRBehavior}) on the strength of the DMI for three values of the external field.}
\label{Prefac}
\end{figure}

The long-range behavior of the total energy $E - 8\pi J$ was found to follow the dependence
\begin{equation}
E -8\pi J= F(A, H,J) \exp\left(-\frac{d}{\delta_{ H}}\right) +{\rm const}, \label{LRBehavior}
\end{equation}
see Fig.\ \ref{LREnergy}. Here $ \delta_{H}= a\sqrt{{J}/{H}}$ is the magnetic length that is independent of $A$. The constant is the correction to the energies of two skyrmions far away from each other due to the contributions of the DMI and the Zeeman interactions.
The prefactor $F$, extracted by the fitting exemplified by Fig.\ \ref{LREnergy}, can be roughly fitted by the formula
\begin{equation}
 F(A, H,J)\approx 60J \left( \frac{A^2}{JH}\right)^2, \label{F_fitted}
\end{equation}
shown by the green line in Fig.\ \ref{Prefac}. The rescaled values of the prefactor are defined as $F\times (0.000225/0.00045)^2$ etc. Within the rigid BP-shape approximation, the DMI and Zeeman energies of a single skyrmion are given by $E_{DM} \propto -|A|\lambda/a$ and $E_Z \propto |H| (\lambda/a)^2$. Minimizing their sum, one obtains $\lambda \propto a|A/H|$, so that Eq. (\ref{F_fitted}) can be written as $F \propto J(\lambda/\delta_H)^2$ that is a plausible result in 2D.

Numerical experiment shows that at a sufficiently small separation $d$ the system undergoes a topological transition from $Q=2$ to $Q=1$. Such a transition cannot be explained within the continuous spin-field model. A similar transition has been seen for skyrmions created in a spin lattice by a magnetic tip \cite{AP-2018}: Nucleation of a $Q = 1$ skyrmion occurred via the creation of a $Q = 0$ skyrmion-antiskyrmion pair with a subsequent collapse of the $Q = -1$ antiskyrmion. In the present case, as the two skyrmions get close to each other, their similar chiralities required by the DMI come in conflict with the field continuity required by the exchange. It becomes then energetically preferable for the two skyrmions to merge into a larger $Q=1$ spin configuration with the favored chirality rather than maintain a frustrated $Q=2$ configuration. At the end it is the discreteness of the lattice that allows for the violation of the conservation of the topological charge.
 \begin{figure}[ht]
\centering
\includegraphics[width=8cm]{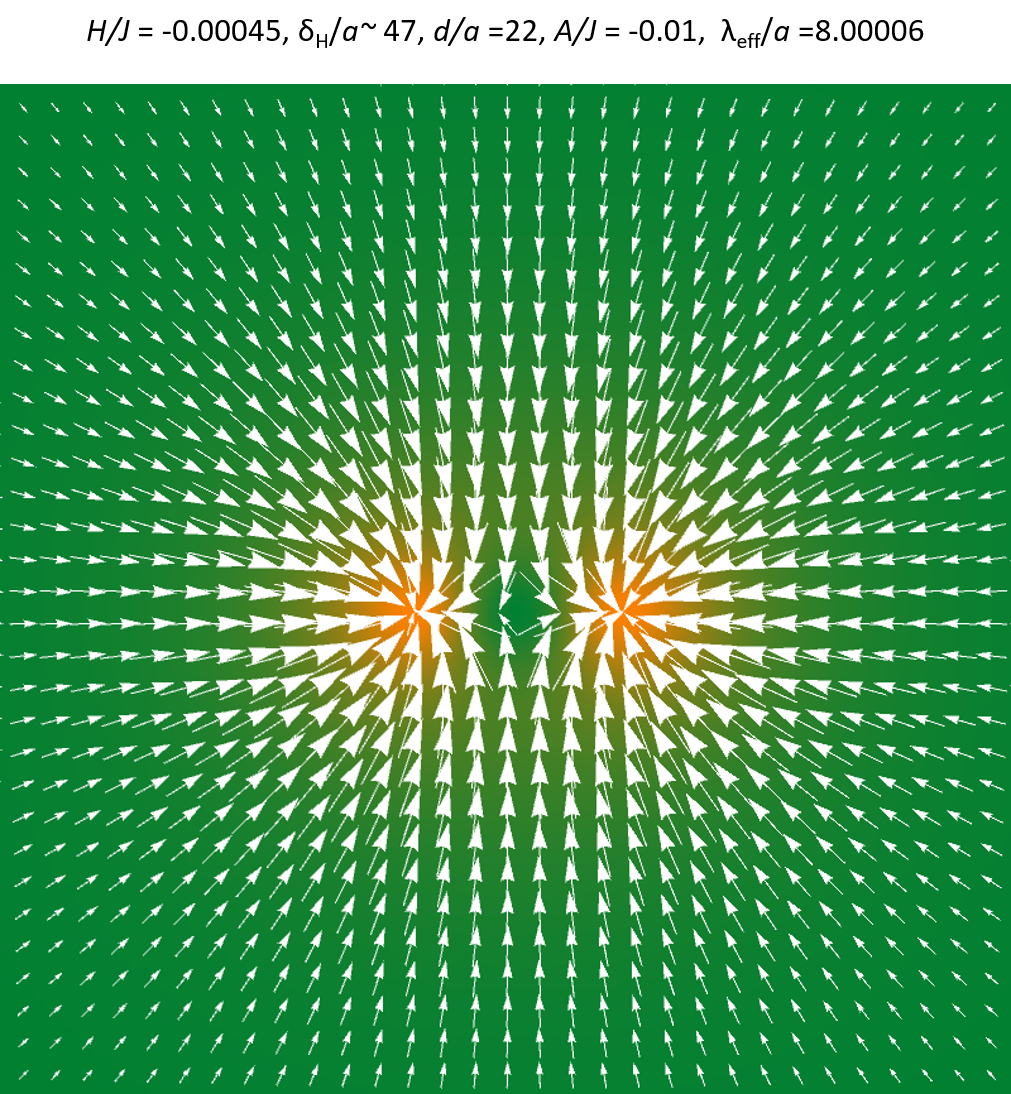}
\caption{Zoomed-in region of the lattice for the transition from $Q=2$ to $Q=1$ immediately before approaching the critical separation $d$. The $Q=2$ configuration is pictured. The white arrows represent the in-plane spin components and the orange regions indicate where the spins are pointing up. }
\label{Merge}
\end{figure}

\begin{figure}[ht]
\hspace{-0.5cm}
\centering
\includegraphics[width=9cm]{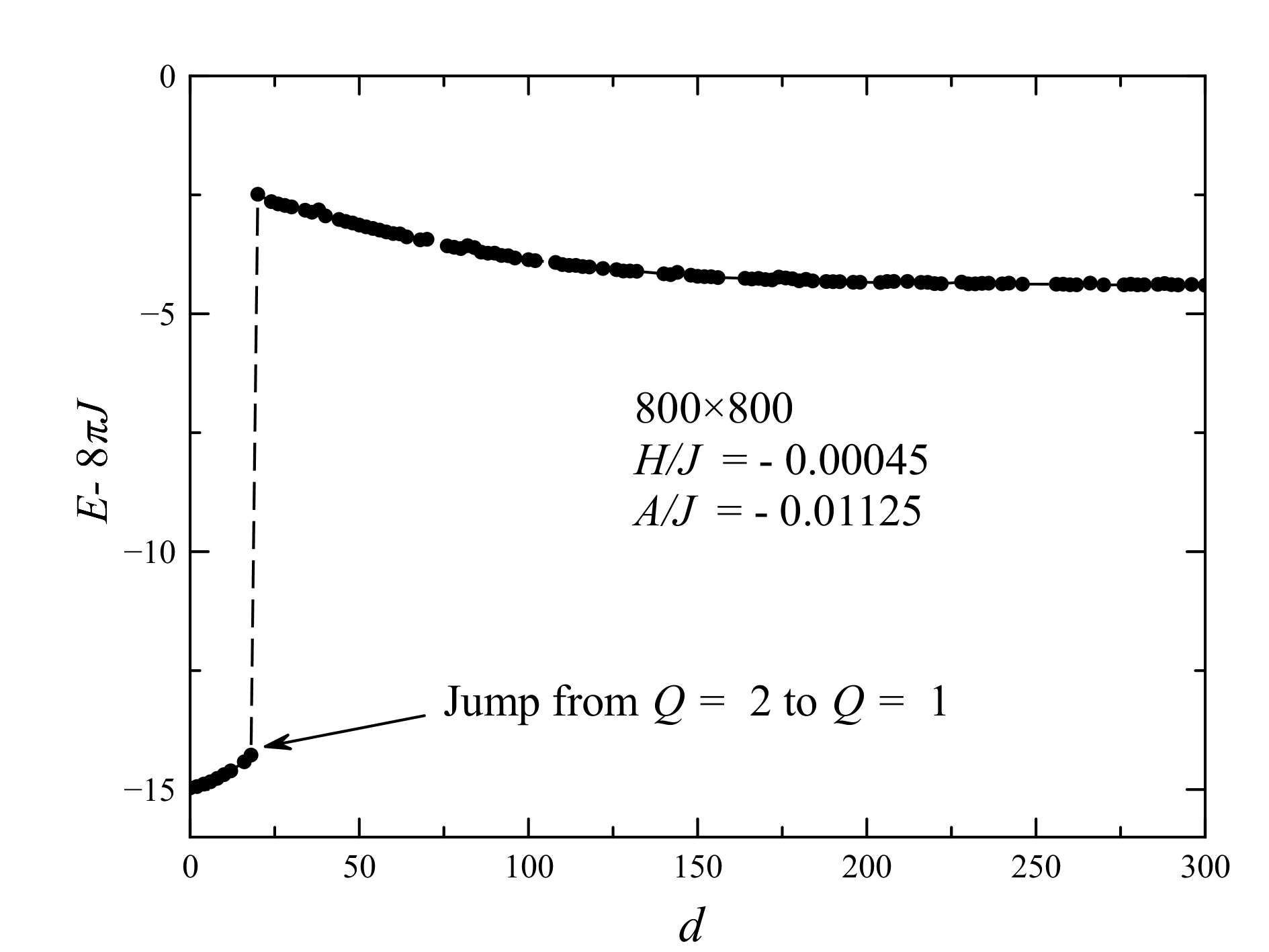}
\caption{Energy of the system in units $E -8\pi J$ with a discontinuity when the system undergoes the transition from $Q=2$ to $Q=1$ at a finite $d$.}
\label{MergeEnergy}
\end{figure}

\section{Dynamics of interacting skyrmions}\label{dynamics}
To study the dynamics of two skyrmions generated by their interaction we solve numerically the system of Landau-Lifshitz equations for the spins on the lattice:
\begin{equation}
\dot{\mathbf{s}}_{i}=\frac{1}{\hbar}\mathbf{s}_{i}\times{\bf H}_{{\rm eff},i}-\frac{\alpha}{\hbar}\mathbf{s}_{i}\times\left(\mathbf{s}_{i}\times\mathbf{H}_{\mathrm{eff},i}\right),\label{LL}
\end{equation}
where $\alpha \ll 1$ is the damping constant. Fourth-order Runge-Kutta ordinary-differential-equation solver with the integration step $0.2$ in the units of $\hbar/J$ has been used. When interactions are weak the dynamics is rather slow, so that the discretization error of the Runge-Kutta method is rather small. Computation was performed on a $300 \times 300$ lattice using $\alpha = 0.1$ and periodic boundary conditions (pbc). Worfram Mathematica with vectorization and compilation has been used on a 20-core Dell Precision Workstation (with 16 cores utilized by Mathematica).

\begin{figure}[ht]
\hspace{-0.5cm}
\centering
\includegraphics[width=9cm]{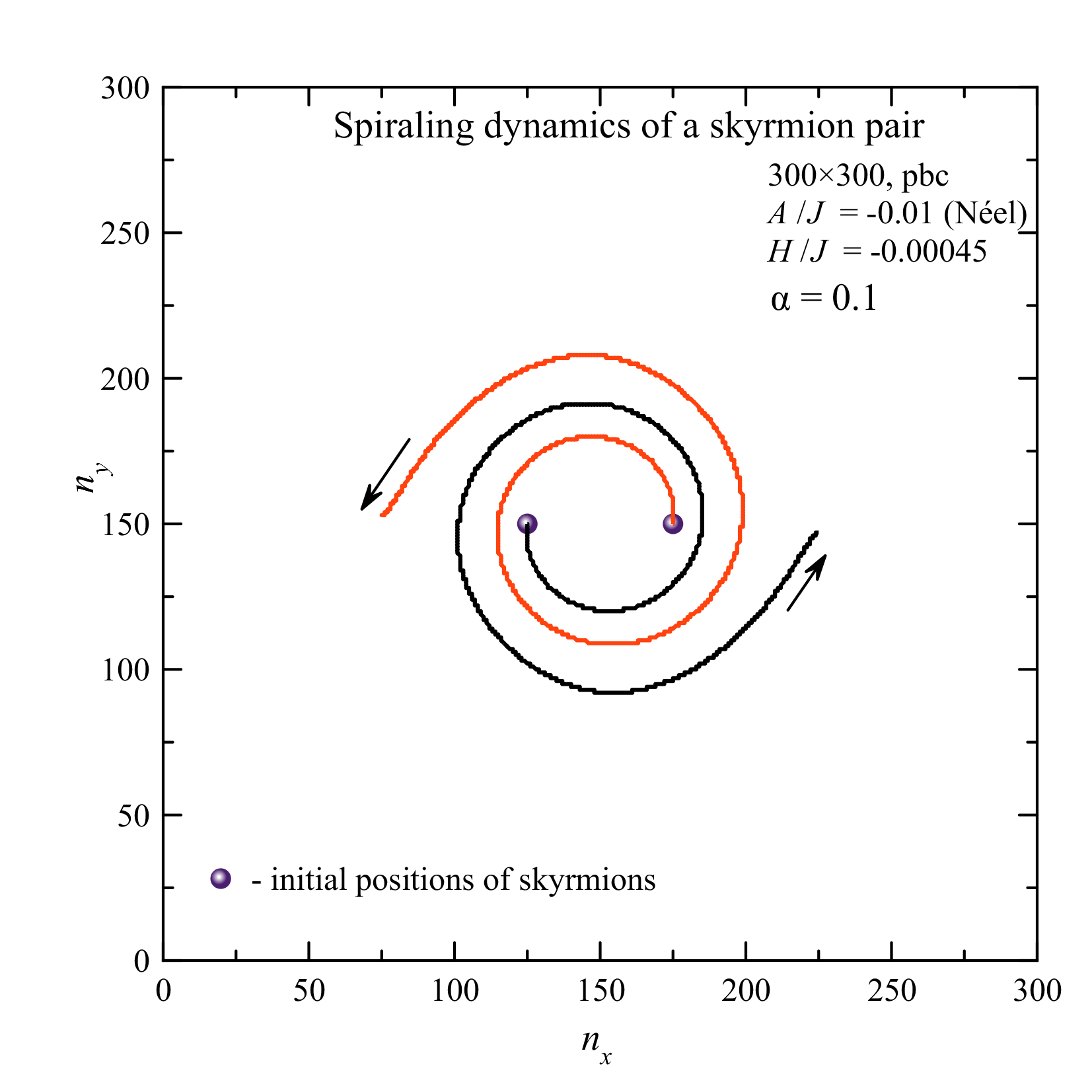}
\caption{Numerically computed trajectories of two repelling skyrmions in the presence of dissipation. Time is measured in units of $\hbar/J$. Skyrmions circle each other counterclockwise, forming the unwinding spiral.}
\label{spiral}
\end{figure}
\begin{figure}[ht]
\hspace{-0.5cm}
\centering
\includegraphics[width=9cm]{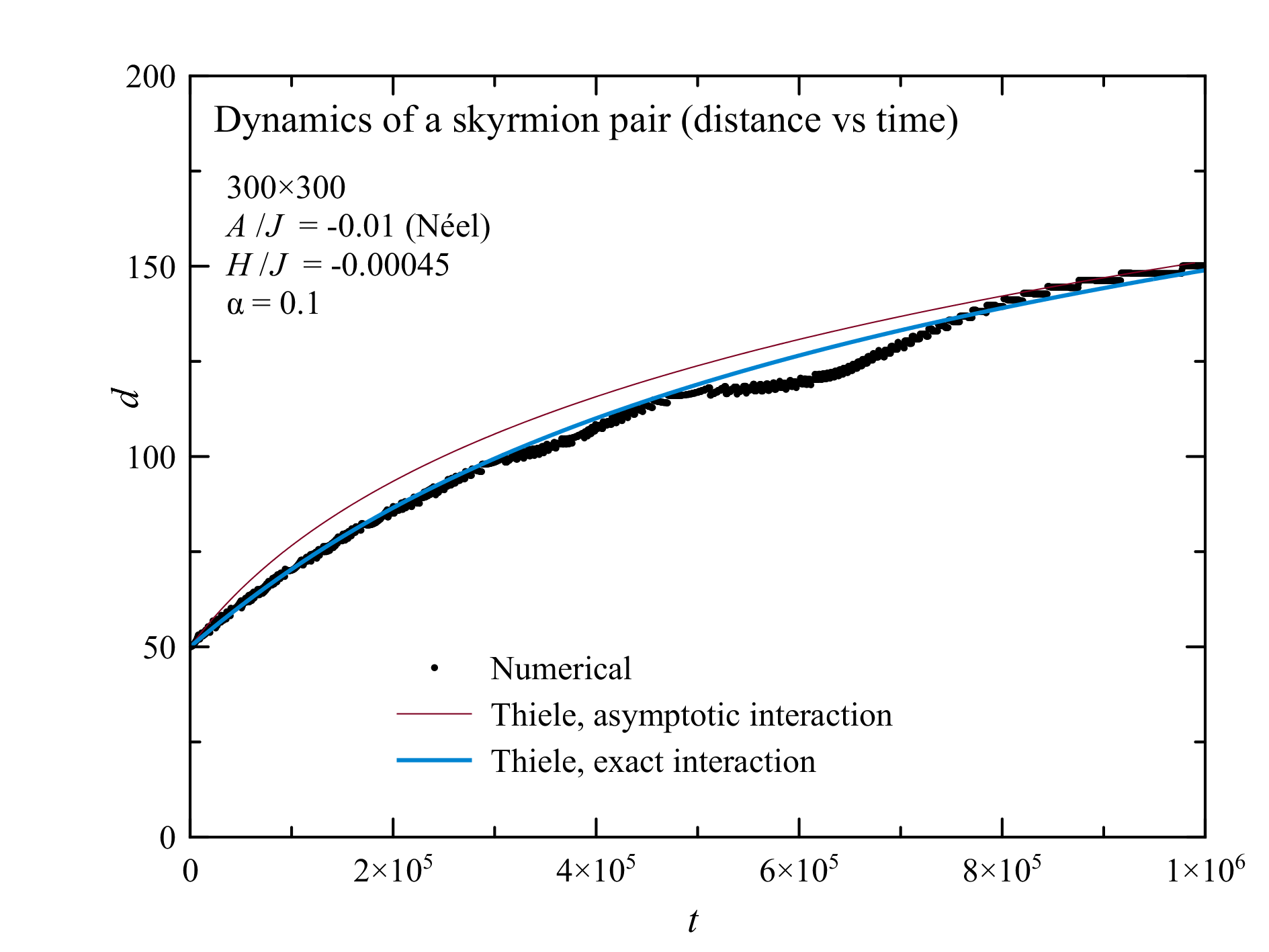}
\caption{Time dependence of the separation of two skyrmions spiraling away from each other as shown in Fig. \ \ref{spiral}. Curves show $d(t)$ computed numerically and obtained with the help of the Thiele equation (see Section \ref{theory}).}
\label{unwinding}
\end{figure}
Numerical results are shown in Figs.\ \ref{spiral} and \ref{unwinding}. In the absence of dissipation, when $\alpha = 0$, (not shown) skyrmions circle each other with their separation remaining constant. The direction of the rotation is always counterclockwise, independently of the sign of the DMI constant $A$. The boundary condition that determines the direction of spins at infinity is then the only factor that can break the symmetry.  Dissipation converts the circular motion into a spiral that unwinds due to the repulsion between the skyrmions. Spiral motion of interacting skyrmions observed in the numerical experiment agrees well with the analytical theory presented in the next section.  This includes numerically obtained time dependence of the separation between the skyrmions. In real experiments such dynamics must freeze as soon as the repulsion force becomes smaller than pinning forces acting on the skyrmions. Bumps in the numerical curve $d(t)$ is a finite-size effect due to the interaction across the boundary in a system with periodic boundary conditions. Due to the latter, skyrmions end up in opposite corners of the system. To the contrast, in the system with free boundary conditions, skyrmions reach the boundaries and disappear there, lowering the system's energy.

\section{Theoretical Arguments Supporting Numerical Results}\label{theory}

Here we argue that the long-range interaction between skyrmions is determined by the field-induced modification of the BP shape. At large distances from the center of the skyrmion, perpendicular components of the skyrmion spin field are small. They satisfy linear equations
\begin{equation}
 -Ja^2 \nabla^2 s_x + |H|s_x  = Ja^2 \nabla^2  s_y - |H|s_y  = 0.
\end{equation}
In the linear approximation only the exchange and Zeeman interactions contribute to these equations, but not the DMI. Writing
\begin{equation}
\nabla^2 = \frac{1}{r}\frac{\partial}{\partial r} r \frac{\partial}{\partial r} + \frac{1}{r^2}\frac{\partial^2}{\partial \phi^2}
\end{equation}
for the 2D Laplacian and choosing
\begin{equation}
s_x = s_{\perp}\cos\phi, \quad s_y = s_{\perp}\sin\phi
\end{equation}
for the N\'{e}el type skyrmion, one obtains the equation for $s_{\perp}$:
\begin{equation}
Ja^2\left(\frac{1}{r}\frac{\partial}{\partial r} r \frac{\partial}{\partial r} - \frac{1}{r^2}\right)s_{\perp}(r) -|H|s_{\perp}(r) = 0
\end{equation}
or, equivalently,
\begin{equation}
\frac{\partial^2s_{\perp}}{\partial r^2} + \frac{1}{r} \frac{\partial s_{\perp}}{\partial r} - \frac{s_{\perp}}{r^2} - \frac{s_{\perp}}{\delta_H^2} = 0.
\end{equation}

Solution of this equation is $s_{\perp}(r) = CK_1(r/\delta_H)$ where $C$ is a constant and $K_1$ is the Macdonald function. It behaves as $1/r$ at $r \ll \delta_H$, which is the BP behavior, but as $\sqrt{\pi \delta_H/(2r)}\exp(-r/\delta_H)$ at $r \gg \delta_H$. This means that the magnetic field makes the skyrmion field fall out exponentially as $\exp(-r/\delta_H)$ at large distances. Consequently, the overlap of the spin fields of two skyrmions separated by the distance $d$ must fall out as
$\exp(-d/\delta_H)$. The long-range interaction between two skyrmions must behave in a similar manner and this is exactly what we see in the numerical experiment. DMI controls the prefactor in this exponential expression and thus drives the skyrmion repulsion.

Note that the above theoretical argument is basically the same as the one presented by Lin et al. \cite{Lin2013} in their Appendix A. We, however, did not use scaled variables that made the conclusion of Ref.\ \onlinecite{Lin2013} somewhat confusing because it hid independence of the characteristic length, that appears in the interaction law, of the DMI.

We now move to the spiral dynamics of interacting skyrmions described in the previous section. The  corresponding equation of motion turns out to be a conventional Thiele equation \cite{Thiele}. We derived it for self-consistency in the Appendix \ref{A1}. Solving Eq.\ (\ref{Thiele-skyrmion_vector}) for the velocity of the skyrmion $\mathbf{V}$ one obtains:
\begin{equation}
\mathbf{V}=\frac{\mathbf{G}\times\mathbf{F}+\Gamma\mathbf{F}}{G^{2}+\Gamma^{2}},\label{Thiele-vector-resolved}
\end{equation}
Here  ${\bf V} = d{\bf R}/dt$, where ${\bf R}$ is the radius-vector pointing towards one skyrmion from the origin of the coordinate frame chosen in the middle of the line connecting the two skyrmions. The force  ${\bf F}$ follows from the interaction energy we computed, see below. The gyrovector ${\bf G}$ and the dissipation factor $\Gamma$ are given by ${\bf G} = (4\pi \hbar Q/a^2)\hat{z}$  and $\Gamma = 4\pi\hbar\alpha\eta/a^2$ where $\alpha \ll 1$ is the damping parameter and $\eta > 1$ accounts for the deformation of the BP shape of the skyrmion (see Appendix).

The first term in the numerator of Eq.\ (\ref{Thiele-vector-resolved}) describes circular motion of one skyrmion of the pair, with the motion of the other skyrmion being mirrored. The second, dissipative, term in Eq.\ (\ref{Thiele-vector-resolved}) converts the circle into the unwinding spiral. Since skyrmions repel, $\mathbf{F}$ is directed along ${\bf r}$, which means that the skyrmions must rotate counterclockwise. This is what we see in the numerical experiment.

In terms of the interaction energy $U(d)=U(2R)$ computed numerically one has $F=-{\partial U}/{\partial d}>0$. According to Eq.\ (\ref{Thiele-vector-resolved}), the motion of two skyrmions away from each other is described by the autonomous equation:
\begin{equation}
\frac{dR}{dt}=\frac{\Gamma}{G^{2}}F(R)
\end{equation}
that can be integrated as
\begin{equation}
\int\frac{dR}{F(R)}=\frac{\Gamma}{G^{2}}t.
\end{equation}
In our case, the interaction decays exponentially at large distances,
thus the distance between the two skyrmions increases logarithmically
with time. Comparison with numerical results in Fig.\ \ref{unwinding} shows an excellent agreement of Thiele dynamics with numerical experiment.

\section{Proposed Experiment on Skyrmion-Skyrmion Interaction}\label{experiment}

\begin{figure}[ht]
\centering
\includegraphics[width=8cm]{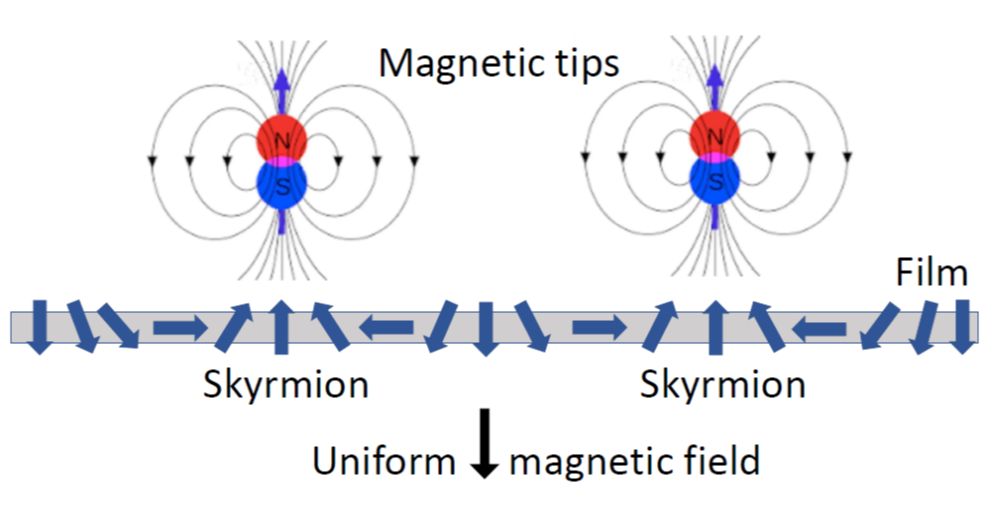}
\caption{Two N\'{e}el-type skyrmions controlled with the MFM tips that we approximate by magnetic dipoles.}
\label{AllMethods}
\end{figure}
 The FCS model is simple theoretically but practically it is impossible to apply a strong magnetic field to a single lattice site. Thus, as an experimental solution, we propose to study the skyrmion-skyrmion interaction using a two-tip magnetic force microscope, see Fig. \ref{AllMethods}. We model magnetic tips by magnetic dipoles. A N\'{e}el-type skyrmion with chirality $\gamma=\pi$ can be supported by a magnetic dipole because the field of the dipole located next to the magnetic film, see Fig.\ \ref{AllMethods} , is similar to the spin configuration of the skyrmion \cite{AP-2018}. Taking $\mathbf{r}=(x,y,h)$ where $h$ is the distance of the tip to the film, with $ r=\sqrt{\rho^2+h^2}$ and $\rho^2= x^2+ y^2$, the components of the induction field of a single dipole are
\begin{eqnarray}
B_{x} & = & -\frac{\mu_{0} m}{4\pi} \frac{3hx}{(\rho^2+h^2)^{5/2}}, \label{TipX} \\
B_{y} & = & -\frac{\mu_{0} m}{4\pi} \frac{3hy}{(\rho^2+h^2)^{5/2}} , \label{TipY} \\
B_{z} & = & \frac{\mu_{0} m}{4\pi} \frac{2h^2-\rho^2}{(\rho^2+h^2)^{5/2}},  \label{TipZ}
\end{eqnarray}
where the in-plane components of the dipole's field favor inward N\'{e}el-type skyrmions and $m$ is the magnetic moment of the tip. The field at the point just below the tip $\mathbf{r}=(0 , 0, h)$ can be used to define the parameters
\begin{equation}
B_{h}= \frac{\mu_{0} m}{2\pi h^3},   \quad H_{h}=g   \mu_{B} S B_{h} ,
\end{equation}
where $H_{h}$ is the corresponding magnetic field just below the tip in the energy units used above, and $S$ is the atomic spin value. This allows one to define the dimensionless ratio $H_{h}/J \ll 1$.

Starting with the same initial state as in Section \ref{Fixed_Spins}, on a lattice of the same size, with $A/J = -0.01$, $H/J= -0.00045$,  we place the two tips at a vertical distance $h = \lambda=15a$ above the centers of the two skyrmions in the biskyrmion and allow the system to relax to the energy minimum, following the same numerical routine described in Section \ref{Fixed_Spins} with an additional Zeeman term to Eq.(\ref{Hamiltonian}),  $-\sum_{i} \mathbf{H}_{i}\cdot \mathbf{s}_{i}$, where $\mathbf{H}_{i}$ is the sum of the two tip fields.  We then move each tip by one lattice spacing in the direction of the separation and compute the next equilibrium. The energy as function of the distance, $d_{\mathrm{tips}}$, between the magnetic tips follows the same behavior as the energy vs $d$ in the fixed spins method. As the tips bring the two skyrmions closer together, their effective size $\lambda$ decreases, leading to the increase in the DMI energy and the energy of the interaction with the tips, while the Zeeman energy due to the stabilizing field decreases.

\begin{figure}[ht]
\hspace{-0.5cm}
\centering
\includegraphics[width=9cm]{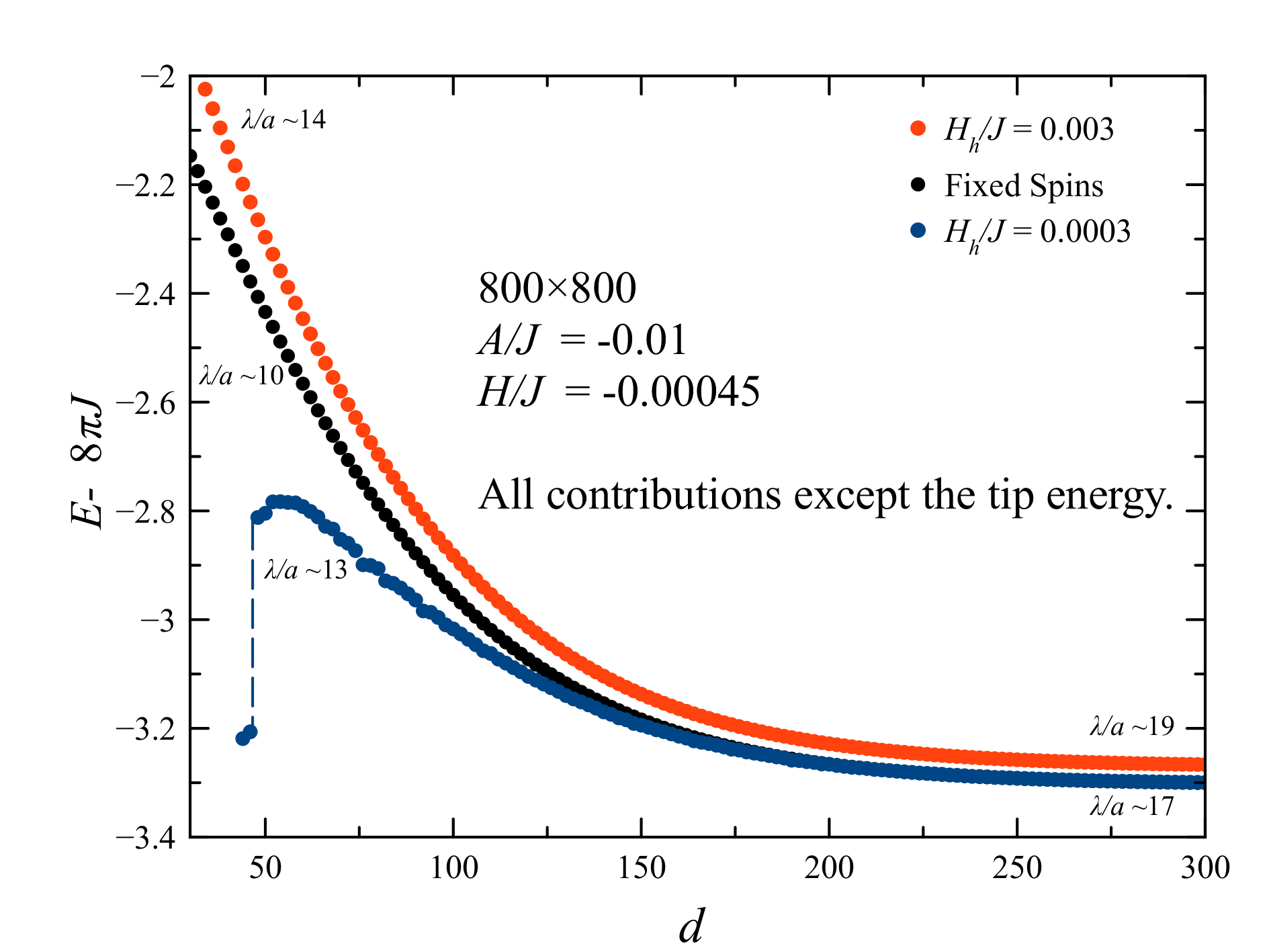}
\caption{Comparison of numerical results for the energy vs separation in the model with the magnetic tips and using the fixed spins approximation. In the model with tips, $d=d_{\mathrm{tips}}$. Good agreement between the two models is observed at sufficiently large separations. }
\label{E-sep}
\end{figure}

\begin{figure}[ht]
\hspace{-0.5cm}
\centering
\includegraphics[width=9cm]{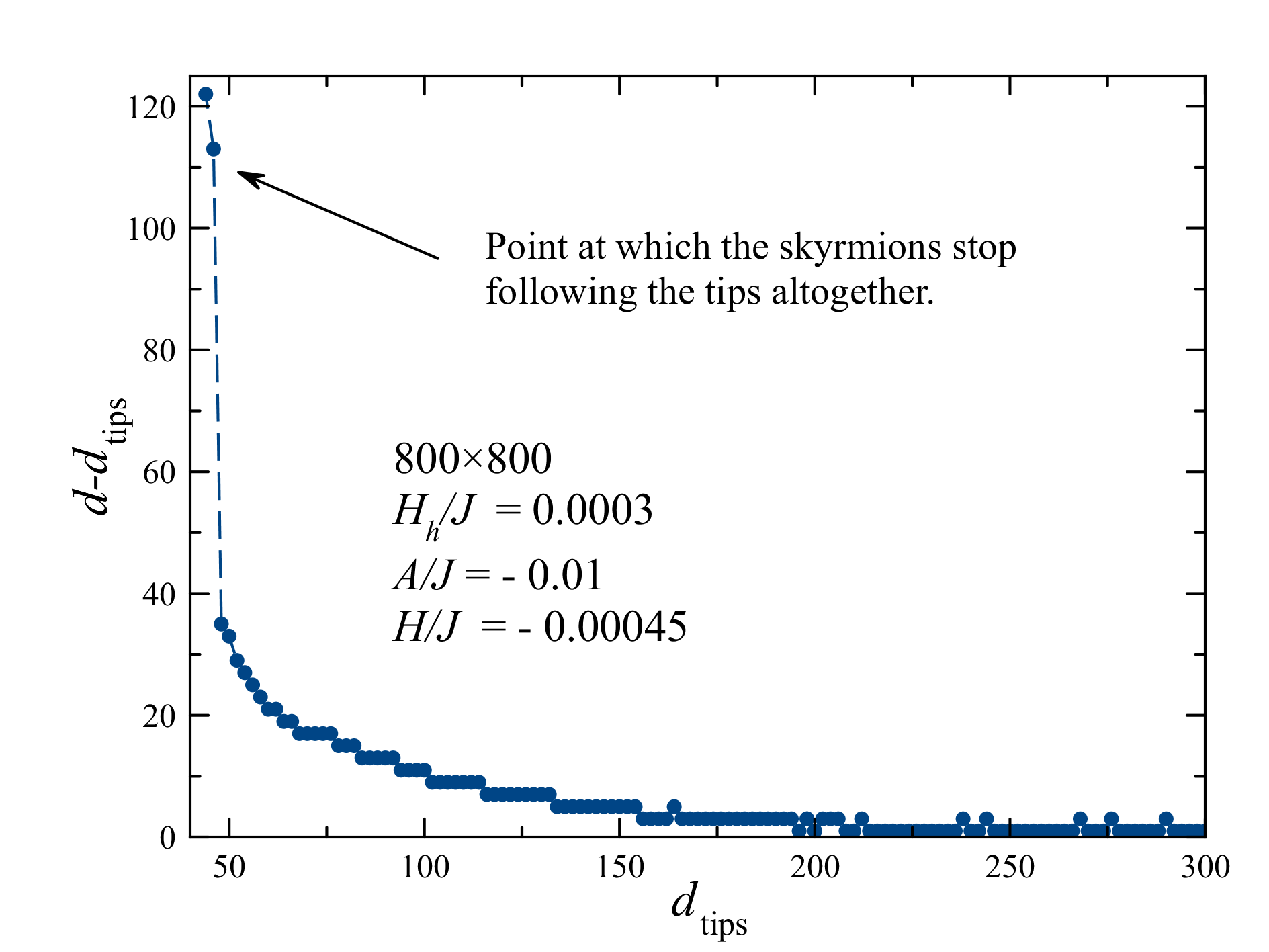}
\caption{The degree to which the skyrmions lag behind the tips in the course of the energy minimization. Here, $d$ refers to the distance between the skyrmions and $d_{\mathrm{tips}}$ is the distance between the magnetic tips. In the plot, $d - d_{\mathrm{tips}}>0$ indicates that the skyrmions are farther apart than the tips. At a critical tip separation, the skyrmion repulsion exceeds their attraction to the tips.}
\label{d}
\end{figure}

\begin{figure}[ht]
\hspace{-0.5cm}
\centering
\includegraphics[width=9cm]{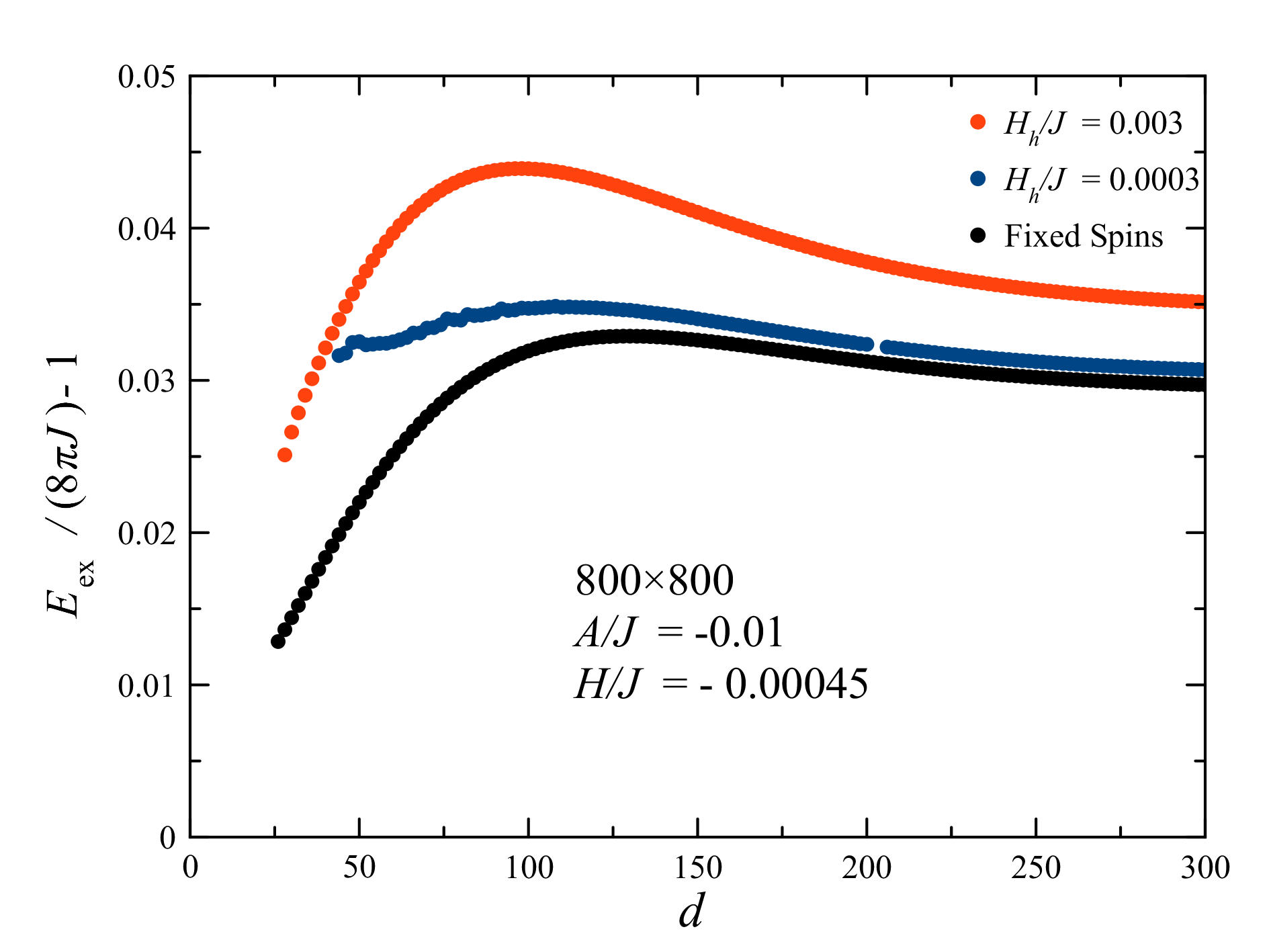}
\caption{The degree of the distortion of the BP skyrmion shape by the magnetic tip and other interactions, measured for two tip strengths by computing the relative difference of the exchange energy from the energy of the BP biskyrmion. For a weak tip, $H_{h}/J = 0.0003$, the distortion is below 3.5\%. For a stronger tip, $H_{h}/J = 0.003$, it is under 4.5\%. The distortion of the BP shape by the interactions that exclude the tip energies found from the FCS model is about 3\%. In the model with tips, $d=d_{\mathrm{tips}}$.}
\label{Eexcomp}
\end{figure}

For the weak tips of strength $H_{h}/J=0.0003 \sim H/J$ the skyrmions lag behind the tips until they cease to follow the tips altogether because the repulsion between the skyrmions exceeds their attraction to the tips, see Fig. \ \ref{d}. Nevertheless, the tips of such a small strength seem well-suited to study the long-range skyrmion-skyrmion interaction because they can hold the skyrmions in position while  not distorting their shape much more than it is already distorted by the other interactions. The energy of such a system computed in our numerical experiment agrees well with the energy computed in the fixed spins model, see Fig.\ \ref{E-sep} . Stronger tips $(H_{h}/J=0.003$) can be used to pin the skyrmion centers  and perhaps can be used to study the short-range interaction, at the cost of more significant deformations of the skyrmion shape. We quantify the deformation as the difference in the exchange energy from $8\pi J$, see Fig.\ \ref{Eexcomp} .

\begin{figure}[ht]
\centering
\includegraphics[width=9cm]{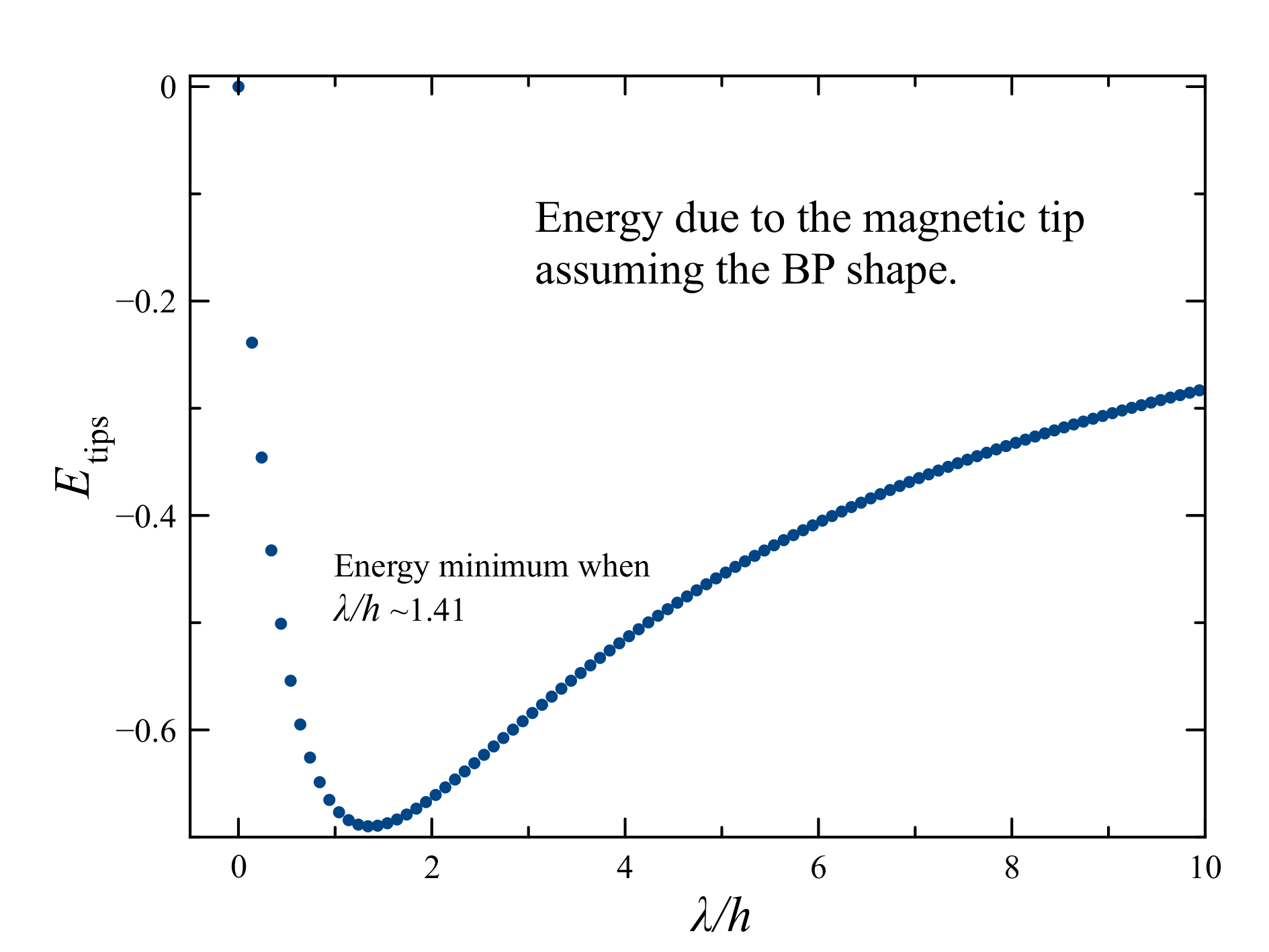}
\caption{Energy of the magnetic tip interaction with a $Q=1$ skyrmion as function of the ratio of the skyrmion size, $\lambda$, to the distance, $h$, of the tip to the film. The minimum occurs at $h \approx 1.4\lambda$. }
\label{Et}
\end{figure}

Weak tips do not lead to strong distortions of the BP shape. Consequently, one can get some analytical predictions for the tip energy that can be used in real experiments. If  $d \gg \lambda$ we can look at the system as comprised of two distinct $Q=1$ objects. Assuming the BP profile, ${\bf s}({\bf r})$, for the $Q=1$ skyrmion, Eq(\ref{BP}) , contribution of the tip field, ${\bf H}({\bf r})$, to the energy can be found by computing the integral $E_{\mathrm{tips}} = - a^{-2}\int dxdy (\mathbf{H} \cdot \mathbf{s})$. The energy due to the tip  is minimized for a skyrmion with in-plane components in the same direction as that of the tip for $\lambda/h = 1.41$  which can be used as a guide for experiments, see Fig.\ \ref{Et} . Choosing the tip distance to the film as $h \sim \lambda$ will allow one to manipulate the skyrmions in the most efficient way.

\section{Discussion} \label{Sec_Discussion}
We have studied skyrmion-skyrmion pair interaction in a minimal model that describes stable skyrmions in a ferromagnetic film. It incorporates ferromagnetic exchange, Dzyaloshinskii-Moriya interaction (DMI), and the external magnetic field. The interest to this problem is two-fold. From purely theoretical perspective it elucidates non-trivial effects arising from topology. In the generic 2D exchange model skyrmions do not interact, the energy of a biskyrmion of topological charge $Q = 2$ is independent of the separation of the two skyrmions, as well as of their sizes. This changes when other interactions are turned on.

From practical perspective, skyrmions have emerged as promising candidates for ultra dense magnetic memory. When experimented with, they often come in close proximity with each other, prompting a natural question how this can affect their stability and motion. Quite generally we find that two skyrmions of the same chirality (that is dictated by the symmetry of the DMI) repel each other with a force that goes down with the separation $d$ as $\exp(-d/\delta_H)$, where $\delta_H$ is the magnetic screening length, independent of the DMI. This is the same result that was obtained by a different method in Ref.\ \onlinecite{Lin2013} but hidden due to their use of rescaled variables. It is confirmed by a rigorous analytical argument. We have shown numerically that the prefactor in this exponential expression depends on the DMI. The latter drives the skyrmion repulsion.

We show that the repulsion between skyrmions of same topological charge acts as a Magnus force rather than the force similar to the Coulomb repulsion between particles of the same electric charge. It generates spiral motion of two skyrmions around each other with separation increasing logarithmically with time due to the damping in the system. The direction of rotation is determined by the boundary conditions for the spins at infinity and is independent of the sign of the DMI. We provide rigorous analytical arguments that explain these findings. In dirty systems one should expect two nearby skyrmions to spiral away from each other until their pair interaction is overcome by pinning. It would be interesting to look for this effect in real experiments.

At small distances the behavior of skyrmion-skyrmion interaction is also repulsive but different from the one described above and more complex. It has been investigated by us numerically. We find that when two skyrmions are pressed towards each other the system undergoes a transition from $Q = 2$ to $Q = 1$, that is, from a biskyrmion to a single $Q = 1$ skyrmion. This happens below some critical separation. Such a transition, prohibited in the continuous spin-field model, occurs entirely due to the discreteness of the atomic lattice. This effect must impose an upper bound on the density of the skyrmion-based magnetic memory.

Experimental study of skyrmion-skyrmion interaction is important if they are to be used for information technology applications. With that in mind we have suggested an experiment that could measure the dependence of the interaction between two skyrmions on their separation. The idea is to hold the skyrmions in positions by magnetic tips and measure interaction between the tips due to the skyrmion-skyrmion repulsion. Although such experiment will require a special MFM design, it may be worth pursuing. By modeling the tips with magnetic dipoles we have shown that the interaction between skyrmions can be measured by this method with good precision.

We also encourage experimentalists to look into simple but striking phenomena predicted in this paper, such as topologically prohibited transitions from $Q = 2$ to $Q = 1$ and the spiral motion of two skyrmions that come close to each other.

\section{Acknowledgements}

This work has been supported by the grant No. DE-FG02-93ER45487 funded
by the U.S. Department of Energy, Office of Science.

 \appendix

\section{Thiele equation for interacting skyrmions}\label{A1}

Dynamics of the magnetic skyrmion satisfies Landau-Lifshitz equation (\ref{LL}). For a weak dissipation, $\alpha \ll 1$, it is equivalent to the Gilbert differential equation for the unit vector of the spin field $\mathbf{s}({\bf r})$:
\begin{equation}
\frac{\partial\mathbf{s}}{\partial t}=\frac{1}{\hbar}\mathbf{s}\times\mathbf{H}_{\mathrm{eff}}(\mathbf{r})-\alpha\mathbf{s}\times\frac{\partial\mathbf{s}}{\partial t}.\label{Gilbert}
\end{equation}

The force acting on the skyrmion is
\begin{equation}
F_{i}=\int\frac{dxdy}{a^2}\mathbf{s}\cdot\partial_{i}\mathbf{H}_{\mathrm{eff}} = -\int\frac{dxdy}{a^2}\mathbf{H}_{\mathrm{eff}}\cdot\partial_{i}\mathbf{s}.\label{Force_def}
\end{equation}
For an isolated skyrmion, this force is non-zero in the presence of additional interactions beyond Eq. (\ref{Hamiltonian}), such as the non-uniform magnetic field or the spin-current term.
To relate the force to the dynamics of the system, it is convenient to re-write Eq.\ (\ref{Gilbert}) as
\begin{equation}
\mathbf{s}\times\left(\mathbf{H}_{\mathrm{eff}}+\hbar\mathbf{s}\times\frac{\partial\mathbf{s}}{\partial t}-\hbar\alpha\frac{\partial\mathbf{s}}{\partial t}\right)=0.
\end{equation}
This means that $\mathbf{s}$ is parallel to the vector in the brackets and, in turn, spatial derivatives of $\mathbf{s}$ are perpendicular
to this vector,
\begin{equation}
\left(\mathbf{H}_{\mathrm{eff}}+\hbar\mathbf{s}\times\frac{\partial\mathbf{s}}{\partial t}-\hbar{\alpha}\frac{\partial\mathbf{s}}{\partial t}\right)\cdot\partial_{i}\mathbf{s}=0.\label{dyn_equil_gradients}
\end{equation}
Integrating this equation over the 2D space and using Eq. (\ref{Force_def}) one obtains
\begin{equation}
F_{i}=\hbar\int\frac{dxdy}{a^2}\left(\mathbf{s}\times\frac{\partial\mathbf{s}}{\partial t}-{\alpha}\frac{\partial\mathbf{s}}{\partial t}\right)\cdot\partial_{i}\mathbf{s}.
\end{equation}

The assumption leading to the Thiele equation is that the skyrmion is moving as a rigid object under the action of the force, that is,
$\mathbf{s}(\mathbf{r},t)=\mathbf{s}(\mathbf{r}-\mathbf{R}(t))\label{structure_rigid}$
and thus
\begin{equation}
\frac{\partial\mathbf{s}}{\partial t}=-V_{j}\partial_{j}\mathbf{s},\qquad\mathbf{V}\equiv\frac{d\mathbf{R}}{dt},\label{s_flow}
\end{equation}
with the summation over repeated indices $i=x,y$. This gives
\begin{equation}
F_{i}=\hbar V_{j}\int\frac{dxdy}{a^2}\left[-\left(\partial_{j}\mathbf{s}\times\partial_{i}\mathbf{s}\right)\cdot\mathbf{s}+\partial_{j}\mathbf{s}\cdot\partial_{i}\mathbf{s}\right]=0,
\end{equation}
which can be written in the form of the Thiele equation:
\begin{equation}
F_{i}=\left(G_{ij}+\Gamma_{ij}\right)V_{j},\label{Thiele-general}
\end{equation}
where
\begin{eqnarray}
G_{ij} & \equiv  & \hbar\int\frac{dxdy}{a^3}\left(\partial_{i}\mathbf{s}\times\partial_{j}\mathbf{s}\right)\cdot\mathbf{s} \\
\Gamma_{ij} &\equiv &\hbar{\alpha}\int\frac{dxdy}{a^3}\partial_{i}\mathbf{s}\cdot\partial_{j}\mathbf{s}.
\end{eqnarray}
Note that $\Gamma_{ij}$ is symmetric while $G_{ij}$ is antisymmetric. Thus, in general,
the first, gyration, term in Eq.\ (\ref{Thiele-general}) can be written in the vector form but the second, dissipation, term cannot.

Below we consider skyrmions that are defined in the $xy$ plane and
rotationally symmetric. Then the components of the gyrotensor are
expressed through the skyrmion's topological charge $Q$ given by Eq.\ (\ref{Q}):
\begin{equation}
G_{xy}=-G_{yx}=G=4\pi\hbar Q/a^{2},\label{Gyrotensor_skyrmion}
\end{equation}
The damping tensor for the skyrmion is diagonal and
related to the exchange energy
\begin{equation}
E_{\mathrm{ex}}=\frac{J}{2}\int d^{2}r\left(\partial_{x}\mathbf{s}\cdot\partial_{x}\mathbf{s}+\partial_{y}\mathbf{s}\cdot\partial_{y}\mathbf{s}\right)
\end{equation}
by
\begin{equation}
\Gamma_{xx}=\Gamma_{yy}=\frac{\alpha\hbar}{a^{2}}\frac{E_{\mathrm{ex}}}{J}.
\end{equation}
For the BP skyrmion one has $E_{\mathrm{ex}}=4\pi J|Q|$. Other interactions (DMI, Zeeman, etc.) deform the BP shape and
increase skyrmion's exchange energy, which can be accounted for by a factor
\begin{equation}
\eta\equiv\frac{E_{\mathrm{ex}}}{4\pi J|Q|}>1.
\end{equation}
The final form of the Thiele equation becomes
\begin{equation}
F_{i}=G_{ij}V_{j}+\Gamma V_{i},\qquad\Gamma=4\pi\hbar\alpha\eta/a^{2}\label{Thiele-skyrmion_components},
\end{equation}
or, in the vector form,
\begin{equation}
\mathbf{F}=-\mathbf{G}\times\mathbf{V}+\Gamma\mathbf{V},\qquad\mathbf{G}=G\hat{z}.\label{Thiele-skyrmion_vector}
\end{equation}

In the model of two interacting skyrmions studied here, there is no force caused by external interactions in Eq. (\ref{Force_def}). Also, the integrals over the area are, in fact, performed over the area enclosing only one of the skyrmions. The interaction force between the skyrmions should be due to their overlap leading to the deformation of the spin field in the region between them. The latter can be most efficiently taken into account phenomenologically via the interaction energy between the skyrmions that is computed numerically in the main part of the paper. Comparison of the solution of the resulting Thiele equation with the full numerical solution for the dynamics of the system of two skyrmions shows that this scheme works very well.

\end{document}